\documentclass[a4paper,11pt]{article}
\usepackage{jheppub}
\usepackage[T1]{fontenc}

\usepackage{slashed}
\usepackage[caption=false]{subfig}
\usepackage{multirow}
\usepackage{feynmf}
\usepackage{tikz}
\usepackage{makecell}

\newcommand\HardNumexample[8]{
\begin{tikzpicture}
  \draw[thick]
      (0,0)node[circle,inner sep=0pt,draw=white,fill=white]{$#1$}
    --(0.7,0)node[circle,inner sep=0pt,draw=white,fill=white]{$#2$}
    --(2*0.7,0)node[circle,inner sep=0pt,draw=white,fill=white]{$#3$}
    --(3*0.7,0)node[circle,inner sep=1pt,draw=white,fill=white]{$#4$}
    --(4*0.7,0)node[circle,inner sep=0pt,draw=white,fill=white]{$#5$};
  \draw[thick]
      (0.7,0)node[circle,inner sep=0pt,draw=white,fill=white]{$#2$}
    --(0.7,0.7)node[circle,inner sep=0pt,draw=white,fill=white]{$#6$};
  \draw[thick]
      (2*0.7,0)node[circle,inner sep=0pt,draw=white,fill=white]{$#3$}
    --(2*0.7,0.7)node[circle,inner sep=0pt,draw=white,fill=white]{$#7$};
  \draw[thick]
      (3*0.7,0)node[circle,inner sep=0pt,draw=white,fill=white]{$#4$}
    --(3*0.7,0.7)node[circle,inner sep=0pt,draw=white,fill=white]{$#8$};
\end{tikzpicture}}

\newcommand\HardNum[8]{
\begin{tikzpicture}
  \draw[thick]
      (0,0)node[circle,inner sep=0pt,draw=white,fill=white]{$#1$}
    --(0.7,0)node[circle,inner sep=0pt,draw=white,fill=white]{$#2$}
    --(2*0.7,0)node[circle,inner sep=0pt,draw=white,fill=white]{$#3$}
    --(3*0.7,0)node[circle,inner sep=1pt,draw=white,fill=white]{$\dots$}
    --(4*0.7,0)node[circle,inner sep=0pt,draw=white,fill=white]{$#4$}
    --(5*0.7,0)node[circle,inner sep=0pt,draw=white,fill=white]{$#5$};
  \draw[thick]
      (0.7,0)node[circle,inner sep=0pt,draw=white,fill=white]{$#2$}
    --(0.7,0.7)node[circle,inner sep=0pt,draw=white,fill=white]{$#6$};
  \draw[thick]
      (2*0.7,0)node[circle,inner sep=0pt,draw=white,fill=white]{$#3$}
    --(2*0.7,0.7)node[circle,inner sep=0pt,draw=white,fill=white]{$#7$};
  \draw[thick]
      (4*0.7,0)node[circle,inner sep=0pt,draw=white,fill=white]{$#4$}
    --(4*0.7,0.7)node[circle,inner sep=0pt,draw=white,fill=white]{$#8$};
\end{tikzpicture}}

\newcommand\HardNumfour[5]{
\begin{tikzpicture}
  \draw[thick]
      (0,0)node[circle,inner sep=0pt,draw=white,fill=white]{$#1$}
    --(0.7,0)node[circle,inner sep=0pt,draw=white,fill=white]{$#2$}
    --(2*0.7,0)node[circle,inner sep=0pt,draw=white,fill=white]{$\cdots$}
    --(3*0.7,0)node[circle,inner sep=0pt,draw=white,fill=white]{$#3$}
    --(4*0.7,0)node[circle,inner sep=0pt,draw=white,fill=white]{$\cdots$}
    --(5*0.7,0)node[circle,inner sep=0pt,draw=white,fill=white]{$#4$}
    --(6*0.7,0)node[circle,inner sep=0pt,draw=white,fill=white]{$#5$};
  \draw[thick]
      (0.7,0)node[circle,inner sep=0pt,draw=white,fill=white]{$#2$}
    --(0.7,0.7)node[circle,inner sep=0pt,draw=white,fill=white]{$\mu_3$};
  \draw[thick]
      (3*0.7-0.4,0.7)node[circle,inner sep=0pt,draw=white,fill=white]{$\mu_k$}
    --(3*0.7,0)node[circle,inner sep=0pt,draw=white,fill=white]{$#3$}
    --(3*0.7+0.4,0.7)node[circle,inner sep=-4.5pt,draw=white,fill=white]{$\mu_{k+1}$};
  \draw[thick]
      (5*0.7,0)node[circle,inner sep=0pt,draw=white,fill=white]{$#4$}
    --(5*0.7,0.7)node[circle,inner sep=0pt,draw=white,fill=white]{$\mu_i$};
\end{tikzpicture}\vspace{-0.2cm}}

\newcommand\Hardtu[6]{
\begin{tikzpicture}
  \draw[thick]
      (0,0)node[circle,inner sep=0pt,draw=white,fill=white]{$#1$}
    --(0.7,0)node[circle,inner sep=0pt,draw=white,fill=white]{$#2$}
    --(1.4,0)node[circle,inner sep=0pt,draw=white,fill=white]{$#3$}
    --(2.1,0)node[circle,inner sep=0pt,draw=white,fill=white]{$#4$};
  \draw[thick]
      (0.7,0)node[circle,inner sep=0pt,draw=white,fill=white]{$#2$}
    --(0.7,0.7)node[circle,inner sep=0pt,draw=white,fill=white]{$#5$};
  \draw[thick]
      (1.4,0)node[circle,inner sep=0pt,draw=white,fill=white]{$#3$}
    --(1.4,0.7)node[circle,inner sep=0pt,draw=white,fill=white]{$#6$};
\end{tikzpicture}}

\newcommand\Hards[4]{
\begin{tikzpicture}
  \draw[thick]
      (0,0)node[circle,inner sep=0pt,draw=white,fill=white]{$#1$}
    --(0.7,0)node[circle,inner sep=0pt,draw=white,fill=white]{$#2$}
    --(1.4,0)node[circle,inner sep=0pt,draw=white,fill=white]{$#3$};
  \draw[thick]
      (0.7,0)node[circle,inner sep=0pt,draw=white,fill=white]{$#2$}
    --(0.7,0.7)node[circle,inner sep=0pt,draw=white,fill=white]{$#4$};
\end{tikzpicture}}

\newcommand\Hardline[2]{
\begin{tikzpicture}
  \draw[thick]
      (0,0)node[circle,inner sep=0pt,draw=white,fill=white] {$#1$}
    --(0.7,0)node[circle,inner sep=0pt,draw=white,fill=white] {$#2$};
\end{tikzpicture}}

\title{\boldmath Massive On-shell Recursion Relations for $n$-point Amplitudes}

\author[a]{Chao Wu}
\author[a,b,c]{and Shou-Hua Zhu}

\affiliation[a]{Department of Physics and State Key Laboratory of Nuclear Physics and Technology,\\Peking University,\\Beijing 100871, China}
\affiliation[b]{Collaborative Innovation Center of Quantum Matter,\\Beijing 100871, China}
\affiliation[c]{Center for High Energy Physics, Peking University,\\Beijing 100871, China}

\emailAdd{wuch7@pku.edu.cn}
\emailAdd{shzhu@pku.edu.cn}

\abstract{We construct two and three-line shifts for tree-level amplitude with massless and/or massive particles, and provide a method to construct general multi-line shifts for all masses. We choose the massless-massive BCFW shift from these shifts and examine its validity in renormalizable theories. Using such a shift, we find that amplitudes with at least one massless vector boson are constructible. This reveals the importance of gauge theory in the construction of amplitudes with massive particles. We also find that this kind of amplitudes have a cancellation related to group structure among different channels, which is essential for constructibility. Furthermore, we show that in the limit of large shift parameter $z$, the amplitude with four massive vector bosons, which can include transverse massive vector particles, have structures proportional to the amplitude with shifted vector particles replaced by Goldstone bosons in the leading order. This is responsible for the failure of massive-massive BCFW recursion relations in the amplitudes with four massive vector bosons.}

\begin{document}

\begin{fmffile}{graph}

\maketitle
\flushbottom

\section{Introduction}
  In comparison with Feynman diagrams, on-shell recursion relations provide a more efficient approach to construct higher-point tree-level amplitudes from lower-point amplitudes. It was first motivated by Britto-Cachazo-Feng-Witten (BCFW) recursion relations~\cite{Britto:2004ap,Britto:2005fq} in the calculation of gluon scattering, and other versions of recursion relations were proposed to study the amplitudes in gauge theories \cite{Badger:2005zh,Risager:2005vk}, gravity theories \cite{Cachazo:2005ca}, supersymmetric theories \cite{Brandhuber:2008pf,Arkani-Hamed:2008owk}, scalar effective field theories \cite{Cheung:2015ota} and more general theories \cite{Cohen:2010mi,Cheung:2015cba}. They based on the same idea, namely using complex deformation of the external momenta and calculating the residues of deformed amplitudes in the complex plane, to collect the information of factorized lower-point amplitudes.

  Contrary to massless particles, the momenta of massive particles cannot be written as a direct product of two spinors. To analyze amplitudes with massive particles, the method of decomposing the massive momenta into two light-like vectors was developed \cite{Schwinn:2005pi,Schwinn:2006ca,Schwinn:2007ee,Craig:2011ws,Boels:2011zz}. However, this formalism is not the most convenient for specific calculation of amplitudes, since it's not little-group covariant. Recently, Arkani-Hamed, Huang and Huang \cite{Arkani-Hamed:2017jhn} introduced a new method by regarding the massive particle as a representation of its little group. In this notation, both amplitudes and complex shifts are simplified into a little-group covariant form. There have been some efforts in constructing massive BCFW shift in the massless-massive case \cite{Aoude:2019tzn,Ballav:2020ese} and massive-massive case \cite{Herderschee:2019dmc,Franken:2019wqr} afterwards.

  After considering little-group invariance, the on-shell constructibility of amplitudes with massive particles was investigated \cite{Franken:2019wqr}. Various multi-line shifts were used to estimate the large-$z$ behavior of amplitudes with all particles massive, where $z$ is the shift parameter. Although people prefer using three and more line shifts to investigate the constructibility, two-line shifts are more convenient in the computation of amplitudes.

  In this work, we take the little-group covariant spinor helicity formalism and regard different spin states from one massive external leg as a whole, so these states should be deformed by a same shift. In the case that all particles are massless, some researches \cite{ArkaniHamed:2008yf,Cheung:2008dn} proved that in a gauge theory coupled to scalars and fermions, any massless amplitude with at least one gluon is two-line constructible, which is a strong conclusion. We want to examine whether the amplitude with both massless and massive particles is also two-line constructible.

  The present paper is organized as follows. In section \ref{sec2} we review the basic idea of recursion relations and construct all possible two and three-line shifts for both massive and massless particles. Assuming the coupling is dimensionless, we evaluate the Feynman rule in a diagrammatic way in section \ref{sec3} and discuss why gauge-fixing is not enough to improve the behavior of amplitude in the large-$z$ limit. In section \ref{sec4} we evaluate $n$-point massive amplitudes with at least one massless vector boson in the large-$z$ limit. We find that these amplitudes vanish in the large-$z$ limit except for the all vector amplitudes, which can have a cancellation among different channels. In section \ref{sec5}, we give the relation between such a cancellation and the group structure of massive vectors, and explore why this cancellation fails in the amplitudes without massless particles. Finally, section \ref{sec6} presents our conclusion and discussion. Appendix \ref{AppA} gives our conventions. Appendix \ref{AppB} explicitly calculates the large-$z$ behavior of polarization vectors in the center-of-mass frame. Appendix \ref{AppC} gives an example of evaluating the diagrammatic expressions.

\section{Recursion relations for all masses}
  \label{sec2}
  For completeness, we review the general complex shift for $A_n$, a tree-level amplitude with $n$ massless particles. For each external particle, we shift their momentum vector $p_i$ by complex-valued vector $r_i$,
  \begin{equation}
    \hat{p}_i^\mu=p_i^\mu+zr_i^\mu\label{momshift},
  \end{equation}
  where $i=1,2,\dots,n$. Now we restrict these shift vectors $r_i$ by three conditions,
  \begin{align}
    \sum_i r_i^\mu=0\label{con1},\\
    r_i\cdot r_j=0\label{con2},\\
    p_i\cdot r_i=0\label{con3}.
  \end{align}

  These three conditions (\ref{con1}--\ref{con3}) respectively guarantee that (a) momentum conservation holds for shifted momenta, (b) shifted momenta are still on-shell, (c) shifted propagators are linear in $z$. We can construct a complex function $\hat{A}_n(z)/z$, whose residue at $z=0$ is the unshifted amplitude $A_n$. Then Cauchy's theorem tells us,
  \begin{equation} \label{masslessCauchy}
    A_n=-\sum_{z_I}\mathrm{Res}_{z=z_I}\frac{\hat{A}_n(z)}{z}+B_n,
  \end{equation}
  where the boundary term $B_n$ is the residue at infinity. The first term on the right-hand side can factorize into two on-shell subamplitudes when the momentum $P_I$ of the internal line goes on-shell,
  \begin{equation} \label{massiveCauchy}
    -\mathrm{Res}_{z=z_I}\frac{\hat{A}_n(z)}{z}=\hat{A}_L(z_I)\frac{1}{P_I^2}\hat{A}_R(z_I).
  \end{equation}

  Now we consider massive amplitudes. The three conditions (\ref{con1}--\ref{con3}) still keep the shifted momenta on-shell $\hat{p}_i^2=p_i^2=m^2$. To generalize formula \eqref{masslessCauchy} into the massive case, we add a new term which corresponds to massive on-shell propagators,
  \begin{equation}
    A_n=-\sum_{z_I}\mathrm{Res}_{z=z_I}\frac{\hat{A}_n(z)}{z}-\sum_{z_J}\mathrm{Res}_{z=z_J}\frac{\hat{A}_n(z)}{z}+B_n.
    \label{rere}
  \end{equation}
  At a $z_J$-pole one of massive internal particles goes on-shell. We use little-group covariant spinors \cite{Arkani-Hamed:2017jhn} to describe massive particles, so the two subamplitudes $\hat{A}_L$ and $\hat{A}_R$ have little-group indices and should be contracted,
  \begin{equation}
    -\mathrm{Res}_{z=z_J}\frac{\hat{A}_n(z)}{z}=\epsilon^{J_1}_{(J'_1}\epsilon^{J_2}_{J'_2}\cdots\epsilon^{J_n}_{J'_n)} \hat{A}_{L,J_1J_2\cdots J_n}(z_J)\frac{1}{P_J^2}\hat{A}_R^{J'_1J'_2\cdots J'_n}(z_J),
  \end{equation}
  where the lower indices in the parenthesis means symmetrization of these indices. Here we only write the little-group indices related to the internal momentum $P_J$ and neglect other little-group indices. If the boundary term $B_n=0$, the $n$-point on-shell amplitude $A_n$ will be completely determined in lower-point on-shell amplitudes and this recursive formula \eqref{rere} becomes an on-shell recursion relation under a valid shift.

  Before we discuss whether amplitudes vanish in the large-$z$ limit, we should construct complex shifts for the composition of all masses. In this section, we will solve equations (\ref{con1}--\ref{con3}) to give all possible shift vectors $r$ in two and three-line shifts. Since massless shifts have been well studied, their generalizations in the massive case will be based on these massless shifts.

  \subsection{Shifting massless particles}
    \label{shiftmassless}
    In spinor-helicity formalism, which is briefly reviewed in appendix \ref{AppA}, we write massless amplitudes in terms of two kinds of Weyl spinor $|p]_\alpha$ and $|p\rangle^{\dot{\alpha}}$. They are two inequivalent fundamental representations of $SL(2,\mathbb{C})$. Both of them can be shifted, we refer to the former as holomorphic shift and the latter as anti-holomorphic shift.

    Since the massive particles takes little-group representation into account, the three conditions (\ref{con1}--\ref{con3}) may not be valid in the massive case. We review some specific massless amplitude recursion relations to translate the three conditions (\ref{con1}--\ref{con3}) into massless spinor-helicity variables. Although all shift-vectors $r_i$ could be non-trivial $r_i\neq0$, two or three-line shifts are enough to construct amplitudes in many applications. Let's start from these few-line shifts.

    \vspace{4pt}
    1) Two-line shift
    \vspace{4pt}

    Since the shifted momentum is linear in $z$, we can't use holomorphic and anti-holomorphic shifts simultaneously for one particle. Otherwise, the momentum conservation condition would be violated. When we shift two external lines $i$ and $j$, there are two choices. We choose holomorphic shift for particle $i$ and anti-holomorphic shift for particle $j$,
    \begin{equation}
      |\hat{i}]=|i]+z|j],\quad
      |\hat{j}\rangle=|j\rangle-z |i\rangle.
      \label{2bcfw}
    \end{equation}
    We call this a $[i,j\rangle$-shift. The shift-vector $r_i=-r_j=r$, so momentum conservation condition \eqref{con1} is automatically satisfied. The shift vector is
    \begin{equation}
      2r^\mu=\langle i|\gamma^\mu|j],
    \end{equation}
    so conditions \eqref{con2} and \eqref{con3} lead to
    \begin{equation}
      2r^2 =\langle ii\rangle[jj],\quad
      2r\cdot p_i =\langle i|p_i|j],\quad
      2r\cdot p_j =\langle i|p_j|j].
    \end{equation}
    We find that $\langle ii\rangle=[jj]=0$ is responsible for the condition \eqref{con2}. Weyl equations $p_j|j]=p_i|i\rangle=0$ are responsible for condition \eqref{con3}.

    \vspace{4pt}
    2) Risager-type three-line shift
    \vspace{4pt}

    In Risager-type, all the shifted external lines are holomorphic shifts. The shifted spinors are
    \begin{equation}
      |\hat{i}]=|i]+z\langle jk\rangle|X],\quad
      |\hat{j}]=|j]+z\langle ki\rangle|X],\quad
      |\hat{k}]=|k]+z\langle ij\rangle|X],
      \label{3risager}
    \end{equation}
    where $|X]$ is an arbitrary reference spinor. Here we ignore the dimension analysis for convenience\footnote{Actually The mass dimension of $\langle jk\rangle$, $\langle ki\rangle$ and $\langle ij\rangle$ is $1$. In order to ensure the dimensionless $z$, we should write the shifts as $|\hat{i}]=|i]+zc\langle jk\rangle|X]$, $|\hat{j}]=|j]+zc\langle ki\rangle|X]$ and $|\hat{k}]=|k]+zc\langle ij\rangle|X]$, where the mass dimension of constant $c$ is $-1$. Since we only want to discuss the large-$z$ behavior, we ignore such constants, which will not change the result of our following analysis.}.
    The shift vectors are
    \begin{equation}
      \begin{aligned}
        2r_i^\mu=\langle jk\rangle\langle i|\gamma^\mu|X],\quad
        2r_j^\mu=\langle ki\rangle\langle j|\gamma^\mu|X],\quad
        2r_k^\mu=\langle ij\rangle\langle k|\gamma^\mu|X].\\
      \end{aligned}
    \end{equation}
    Using Schouten identity $\langle ki\rangle\langle j|+\langle ki\rangle\langle j|+\langle ij\rangle\langle k|=0$, we easily verify condition \eqref{con1} $r_i+r_j+r_k=0$. We find that $[XX]=0$ ensures condition \eqref{con2}, and that Weyl equations $p_i|i\rangle=p_j|j\rangle=p_k|k\rangle=0$ are responsible for condition \eqref{con3}.

    \vspace{4pt}
    3) BCFW-type three-line shift
    \vspace{4pt}

    In BCFW-type, only one shifted external line is anti-holomorphic shift, the other shifted external lines are holomorphic shifts. The shifted spinors are,
    \begin{equation}
      |\hat{i}]=|i]+z\langle jX\rangle|k],\quad
      |\hat{j}]=|j]+z\langle Xi\rangle|k],\quad
      |\hat{k}\rangle=|k\rangle+z\langle ij\rangle|X\rangle,
      \label{3bcfw}
    \end{equation}
    where $|X\rangle$ is an arbitrary reference spinor. The shift vectors are
    \begin{equation}
      \begin{aligned}
        2r_i^\mu=\langle jX\rangle\langle i|\gamma^\mu|k],\quad
        2r_j^\mu=\langle Xi\rangle\langle j|\gamma^\mu|k],\quad
        2r_k^\mu=\langle ij\rangle\langle X|\gamma^\mu|k].\\
      \end{aligned}
    \end{equation}
    We use Schouten identity to verify condition \eqref{con1}, $\sum_ir_i=0$. We find that $[kk]=0$ ensures condition \eqref{con2} and Weyl equations $p_i|i\rangle=p_j|j\rangle=p_k|k]=0$ are responsible for condition \eqref{con3}.

    So far all possible two and three-line shifts in the massless case are presented. Since Risager and BCFW-type shifts are the only two distinct classes of recursion relations \cite{Cheung:2015cba}, the analysis of four and more-line shifts are just generalizations of three-line shifts. They are sufficient for analyzing mixed or massive recursion relations. We find that the antisymmetry of brackets $\langle\cdots\rangle$ and $[\cdots]$ ensure condition \eqref{con2}, and the equation of motion for massless particles is responsible for condition \eqref{con3}.

  \subsection{Shifting mixed particles}

    Now let's think about how to satisfy the three conditions when massive particles are taken into account. The massive spinor-helicity variables $|\mathbf{p}^I]_\alpha$ and $|\mathbf{p}^I\rangle^{\dot{\alpha}}$ are not only fundamental representations of Lorentz group $SO(3,1)$, but also fundamental representations of little group $SU(2)_{LG}$. We write them as $(\frac{1}{2};\frac{1}{2},0)$ and $(\frac{1}{2};0,\frac{1}{2})$, which are the representations of $SU(2)_{LG}\times SO(3,1)$. In the massive case, the antisymmetric brackets
    \begin{equation}
      [\mathbf{i}^I\mathbf{i}^{I'}]=-\langle\mathbf{i}^I\mathbf{i}^{I'}\rangle=m_i^2\epsilon^{II'}\neq 0
    \end{equation}
    and massive Dirac equation
    \begin{equation}
      p|\mathbf{p}^I]=-m|\mathbf{p}^I\rangle\neq 0
    \end{equation}
    are different from the massless case. Therefore, conditions \eqref{con2} and \eqref{con3} wouldn't be satisfied automatically in the massive case. For example, we can give up condition \eqref{con3} for massive particles and construct a shift, where masses are no longer invariants. The validity of this kind of shifts was examined numerically for the amplitudes with graviton and scalar bosons \cite{Britto:2021pud}. Here we try to satisfy all conditions.

    The key point is to use some variables to contract the little-group index or Weyl-spinor index of massive spinors. For example, we can introduce an unknown variable $\eta^I$, which is the representation $(\frac{1}{2};0,0)$. We use $\eta^I$ to contract the little-group index of $|\mathbf{i}^I]$, so the inner product $\eta_I|\mathbf{i}^I]$ gives the antisymmetric bracket
    \begin{equation}
      \eta_I[\mathbf{i}^I\mathbf{i}^{I'}]\eta_{I'}=m_i^2\eta_I\epsilon^{II'}\eta_{I'}=0,
    \end{equation}
    which ensures condition \eqref{con2}. Another way is to introduce $|Y]$ and contract the Weyl-spinor index of $|\mathbf{i}^I]$, which has been used in ref. \cite{Franken:2019wqr}.

    Our strategy is the former one. First, we take massless shifts in Section \ref{shiftmassless} and replace the massless variables $|p]_\alpha$ and $|p\rangle^{\dot{\alpha}}$ with the massive spinor-helicity variables $|\mathbf{p}^I]_\alpha$ and $|\mathbf{p}^I\rangle^{\dot{\alpha}}$ for massive particles, while the spinor part of the shifts remain as massless shifts. Then we introduce some unknown variables with little-group indices ($\eta^I$, $\zeta^J$, $\xi^L$, etc.), whose number equals the number of massive shifted external legs. With the shifts which have been replaced multiplied by or contracted with these unknown variables, condition \eqref{con2} must be satisfied. Since the above manipulation is based on massless shifts, condition \eqref{con1} is still satisfied. Finally, we use the last condition \eqref{con3} to determine these unknown variables.

    \subsubsection{Two-line shifts for mixed particles}

      Let's consider two-line shifts for massive particle $i$ and massless particle $j$. There are two ways to shift them. We can choose the $[i,\mathbf{j}\rangle$-shift: the massless line $i$ is shifted holomorphically, while the massive line $j$ is shifted anti-holomorphically. We introduce one unknown $\zeta^J$, so the shifted spinors are
      \begin{equation}
        |\hat{i}]=|i]+z|\mathbf{j}^J]\zeta_J,\quad
        |\mathbf{\hat{j}}^J\rangle=|\mathbf{j}^J\rangle-z|i\rangle\zeta^J.
        \label{ij1}
      \end{equation}

      The shift vector is $2r^\mu=\langle i|\gamma^\mu|\mathbf{j}^J]\zeta_J$. It is orthogonal to the massless momentum $p_i$ because of the Weyl equation $\langle i|p_i=0$.  Condition \eqref{con3} leads to
      \begin{equation}
        2p_j\cdot r=\langle i|p_j|\mathbf{j}^J]\zeta_J=m_j\langle i\mathbf{j}^J\rangle\zeta_J=0.
      \end{equation}
      The solution is $\zeta^J=\langle i\mathbf{j}^J\rangle$. We substitute the solution into eq.\,\eqref{ij1}, so the explicit form of  $[i,\mathbf{j}\rangle$-shift is
      \begin{equation}
        |\hat{i}]=|i]+z|\mathbf{j}^J]\langle i\mathbf{j}_J\rangle,\quad
        |\mathbf{\hat{j}}^J\rangle=|\mathbf{j}^J\rangle-z|i\rangle\langle i\mathbf{j}^J\rangle.
        \label{ij2}
      \end{equation}

      Another choice is the $[\mathbf{j},i\rangle$-shift: the massless line $i$ uses holomorphic shift, the massive line $j$ uses anti-holomorphic shift. For real-valued momenta, angle and square spinors are not independent. For massless particles, we have $(|i\rangle)^*=[i|$ and $(|i])^*=\langle i|$. For the massive particle $j$, the complex conjugation of massive spinors lowers the little indices: $(|\mathbf{j}^J\rangle)^*=[\mathbf{j}_J|$ and $(|\mathbf{j}^J])^*=\langle\mathbf{j}_J|$. Therefore, the $[\mathbf{j},i\rangle$-shift can be implemented from the complex conjugate of $[i,\mathbf{j}\rangle$-shift,
      \begin{equation}
        |\mathbf{\hat{j}}^J]=|\mathbf{j}^J]+z|i][i\mathbf{j}^J],\quad
        |\hat{i}\rangle=|i\rangle-z |\mathbf{j}^J\rangle[i\mathbf{j}_J].
      \end{equation}

    \subsubsection{Three-line Risager type shifts for mixed Particles}
      We take massless shift \eqref{3risager} and choose one or two shifted particles to be massive. Since all shifts are holomorphic shifts in Risager type, each shift vector must be $r^\mu\propto \langle\cdots|\gamma^\mu|X]$. Condition \eqref{con2} is satisfied, because $r_i\cdot r_j\propto[XX]=0$.

      \vspace{4pt}
      1) One massive and two massless
      \vspace{4pt}

      Let particle $i$ be massive particle and particles $j$ and $k$ be massless particles. We introduce one unknown $\eta^I$, so the shifted spinors are
      \begin{equation}
        |\mathbf{\hat{i}}^I]=|\mathbf{i}^I]+z\eta^I\langle jk\rangle|X],\quad
        |\hat{j}]=|j]+z\eta^I\langle k\mathbf{i}_I\rangle|X], \quad
        |\hat{k}]=|k]+z\eta^I\langle \mathbf{i}_Ij\rangle|X].
      \end{equation}
      The shift vectors are
      \begin{equation}
        2r_i^\mu=\eta^I\langle jk\rangle\langle\mathbf{i}_I|\gamma^\mu|X],\quad
        2r_j^\mu=\eta^I\langle k\mathbf{i}_I\rangle\langle j|\gamma^\mu|X],\quad
        2r_k^\mu=\eta^I\langle \mathbf{i}_I j\rangle\langle k|\gamma^\mu|X],
      \end{equation}
      so condition \eqref{con3} leads to
      \begin{equation}
        2p_i\cdot r_i=\eta^I\langle jk\rangle\langle\mathbf{i}_I|p_i|X]= m_i\eta^I\langle jk\rangle[\mathbf{i}_IX]=0.
      \end{equation}

      The solution is $\eta^I=[\mathbf{i}^IX]$.

      \vspace{4pt}
      2) Two massive and one massless
      \vspace{4pt}

      Let particles $i$ and $j$ be massive particles and particle $k$ be massless particle. We introduce two unknowns $\eta^I$ and $\zeta^J$, so the shifted spinors are
      \begin{equation} \begin{aligned}
        |\mathbf{\hat{i}}^I]&=|\mathbf{i}^I]+z\eta^I\zeta^J\langle \mathbf{j}_Jk\rangle|X],\\
        |\mathbf{\hat{j}}^J]&=|\mathbf{j}^J]+z\eta^I\zeta^J\langle k\mathbf{i}_I\rangle|X],\\
        |\hat{k}]&=|k]+z\eta^I\zeta^J\langle \mathbf{i}_I\mathbf{j}_J\rangle|X].
      \end{aligned} \end{equation}
      The shift vectors are
      \begin{equation} \begin{aligned}
        2r_i^\mu&=\eta^I\zeta^J\langle \mathbf{j}_Jk\rangle\langle\mathbf{i}_I|\gamma^\mu|X],\\
        2r_j^\mu&=\eta^I\zeta^J\langle k\mathbf{i}_I\rangle\langle\mathbf{j}_J|\gamma^\mu|X],\\
        2r_k^\mu&=\eta^I\zeta^J\langle \mathbf{i}_I\mathbf{j}_J\rangle\langle k|\gamma^\mu|X],
      \end{aligned} \end{equation}
      so condition \eqref{con3} leads to
      \begin{equation}
        \begin{aligned}
          2p_i\cdot r_i &=\eta^I\zeta^J\langle \mathbf{j}_Jk\rangle\langle\mathbf{i}_I|p_i|X]= m_i\eta^I\zeta^J\langle \mathbf{j}_Jk\rangle[\mathbf{i}_IX]=0,\\
          2p_j\cdot r_j &=\eta^I\zeta^J\langle k\mathbf{i}_I\rangle\langle\mathbf{j}_J|p_j|X]= m_j\eta^I\zeta^J\langle k\mathbf{i}_I\rangle[\mathbf{j}_JX]=0.\\
        \end{aligned}
      \end{equation}
      The shift vectors should be non-trivial, so $\zeta^J\langle \mathbf{j}_Jk\rangle\neq0$, $\eta^I\zeta^J\langle k\mathbf{i}_I\rangle\neq0$. The only solutions are $\eta^I=[\mathbf{i}^IX]$, $\zeta^J=[\mathbf{j}^JX]$.

    \subsubsection{Three-line BCFW type shifts for mixed particles}
    We take massless shift \eqref{3bcfw} and choose one or two shifted particles to be massive. If particle $k$ is massive, each shift vector in this type of shifts must be $r^\mu\propto \langle\cdots|\gamma^\mu|\mathbf{k}^K]\xi_K$. Condition \eqref{con2} is satisfied, because $r_i\cdot r_j\propto[\mathbf{k}^K\mathbf{k}^{K'}]\xi_K\xi_{K'}=\epsilon^{KK'}\xi_K\xi_{K'}=0$. If particle $k$ is massless, condition \eqref{con2} is also satisfied for the same reason as the case of massless three-line BCFW type shifts.

      \vspace{4pt}
      1) One massive and two massless
      \vspace{4pt}

      Since BCFW type recursion relations use holomorphic and anti-holomorphic shift, there are two kinds of compositions. The first kind is that the massive particle uses holomorphic shift. Let particle $i$ be massive particle and particles $j$ and $k$ be massless particle. We introduce one unknown $\eta^I$, so the shifted spinors are
      \begin{equation}
        |\mathbf{\hat{i}}^I]=|\mathbf{i}^I]+z\eta^I\langle jX\rangle|k],\quad
        |\hat{j}]=|j]+z\eta^I\langle X\mathbf{i}_I\rangle|k], \quad
        |\hat{k}\rangle=|k\rangle+z\eta^I\langle\mathbf{i}_Ij\rangle|X\rangle.
      \end{equation}
      The shift vectors are
      \begin{equation}
        2r_i^\mu=\eta^I\langle jX\rangle\langle\mathbf{i}_I|\gamma^\mu|k],\quad
        2r_j^\mu=\eta^I\langle X\mathbf{i}_I\rangle\langle j|\gamma^\mu|k],\quad
        2r_k^\mu=\eta^I\langle \mathbf{i}_I j\rangle\langle X|\gamma^\mu|k],
      \end{equation}
      so condition \eqref{con3} leads to
      \begin{equation}
        2p_i\cdot r_i=\eta^I\langle Xj\rangle\langle\mathbf{i}_I|p_i|k]= m_i\eta^I\langle Xj\rangle[\mathbf{i}_Ik]=0.
      \end{equation}
      The solution is $\eta^I=[\mathbf{i}^Ik]$.

      The second kind is that the massive particle uses anti-holomorphic shift. Let particle $k$ be massive particle and particles $i$ and $j$ be massless particles. We introduce one unknown $\xi^K$, so the shifted spinors are
      \begin{equation}
        |\hat{i}]=|i]+z\xi_K\langle jX\rangle|\mathbf{k}^K],\quad
        |\hat{j}]=|j]+z\xi_K\langle Xi\rangle|\mathbf{k}^K],\quad
        |\mathbf{\hat{k}}^K\rangle=|\mathbf{k}^K\rangle-z\xi^K\langle ij\rangle|X\rangle.
      \end{equation}
      The shift vectors are
      \begin{equation}
        2r_i^\mu=\xi_K\langle jX\rangle\langle i|\gamma^\mu|\mathbf{k}^K],\quad
        2r_j^\mu=\xi_K\langle Xi\rangle\langle j|\gamma^\mu|\mathbf{k}^K],\quad
        2r_k^\mu=\xi_K\langle ij\rangle\langle X|\gamma^\mu|\mathbf{k}^K],
      \end{equation}
      so condition \eqref{con3} leads to
      \begin{equation}
        2p_k\cdot r_k=\xi_K\langle ij\rangle\langle X|p_k|\mathbf{k}^K]=-m_k\xi_K\langle ij\rangle\langle X\mathbf{k}^K\rangle=0.
      \end{equation}
      The solution is $\xi^K=\langle X\mathbf{k}^K\rangle$.

      \vspace{4pt}
      2) Two massive and one massless
      \vspace{4pt}

      Let particles $i$ and $j$ be massive particles and particle $k$ be massless particle. We introduce two unknowns $\eta^I$ and $\zeta^J$,  so the shifted spinors are
      \begin{equation} \begin{aligned}
        |\mathbf{\hat{i}}^I]&=|\mathbf{i}^I]+z\eta^I\zeta^J\langle\mathbf{j}_JX\rangle|k]\\
        |\mathbf{\hat{j}}^J]&=|\mathbf{j}^J]+z\eta^I\zeta^J\langle X\mathbf{i}_I\rangle|k]\\
        |\hat{k}\rangle&=|k\rangle+z\eta^I\zeta^J\langle \mathbf{i}_I\mathbf{j}_J\rangle|X\rangle.
      \end{aligned} \end{equation}
      The shift vectors are
      \begin{equation} \begin{aligned}
        2r_i^\mu=\eta^I\zeta^J\langle \mathbf{j}_JX\rangle\langle\mathbf{i}_I|\gamma^\mu|k],\\
        2r_j^\mu=\eta^I\zeta^J\langle X\mathbf{i}_I\rangle\langle\mathbf{j}_J|\gamma^\mu|k],\\
        2r_k^\mu=\eta^I\zeta^J\langle \mathbf{i}_I\mathbf{j}_J\rangle\langle X|\gamma^\mu|k],
      \end{aligned} \end{equation}
      so condition \eqref{con3} leads to
      \begin{equation}
        \begin{aligned}
          2p_i\cdot r_i &=\eta^I\zeta^J\langle \mathbf{j}_JX\rangle\langle\mathbf{i}_I|p_i|k]= m_i\eta^I\zeta^J\langle \mathbf{j}_JX\rangle[\mathbf{i}_Ik]=0,\\
          2p_j\cdot r_j &=\eta^I\zeta^J\langle X\mathbf{i}_I\rangle\langle\mathbf{j}_J|p_j|k]= m_j\eta^I\zeta^J\langle X\mathbf{i}_I\rangle[\mathbf{j}_Jk]=0.
        \end{aligned}
      \end{equation}
      The solutions are $\eta^I=[\mathbf{i}^Ik]$, $\zeta^J=[\mathbf{j}^Jk]$.

      Let particles $i$ and $k$ be massive particles and particle $j$ be massless particle. We introduce two unknowns $\eta^I$ and $\xi^K$, so the shifted spinors are
      \begin{equation} \begin{aligned}
        |\mathbf{\hat{i}}^I]&=|\mathbf{i}^I]+z\eta^I\xi^K\langle jX\rangle|\mathbf{k}_K],\\
        |\hat{j}]&=|j]+z\eta^I\xi^K\langle X\mathbf{i}_I\rangle|\mathbf{k}_K],\\
        |\mathbf{\hat{k}}^K\rangle&=|\mathbf{k}^K\rangle-z\eta^I\xi^K\langle \mathbf{i}_I j\rangle|X\rangle.
      \end{aligned} \end{equation}
      The shift vectors are
      \begin{equation} \begin{aligned}
        2r_i^\mu&=\eta^I\xi^K\langle jX\rangle\langle\mathbf{i}_I|\gamma^\mu|\mathbf{k}_K],\\
        2r_j^\mu&=\eta^I\xi^K\langle X\mathbf{i}_I\rangle\langle j|\gamma^\mu|\mathbf{k}_K],\\
        2r_k^\mu&=\eta^I\xi^K\langle \mathbf{i}_Ij\rangle\langle X|\gamma^\mu|\mathbf{k}_K],
      \end{aligned} \end{equation}
      so condition \eqref{con3} leads to
      \begin{equation}
        \begin{aligned}
          2p_i\cdot r_i &=\eta^I\xi^K\langle jX\rangle\langle\mathbf{i}_I|p_i|\mathbf{k}_K]= m_i\eta^I\xi^K\langle jX\rangle[\mathbf{i}_I\mathbf{k}_K]=0,\\
          2p_k\cdot r_k &=\eta^I\xi^K\langle \mathbf{i}_Ij\rangle\langle X|p_k|\mathbf{k}_K]= m_j\eta^I\xi^K\langle \mathbf{i}_Ij\rangle\langle X\mathbf{k}_K\rangle=0.
        \end{aligned}
      \end{equation}
      The solutions are $\eta^I=[\mathbf{i}^I|p_k|X\rangle$, $\xi^K=[\mathbf{k}^KX]$. The solutions become more complicated.

  \subsection{Shifting massive particles}
    The all-massive recursion relations have been worked out to study the constructibility of all-massive amplitudes in spontaneously broken gauge theories \cite{Franken:2019wqr}. Now we want to reproduce these massive shifts with our method. Similarly, we introduce $n$ unknown variables for $n$-line shifts and then solve eq.\,\eqref{con3} for these unknown variables.

    \subsubsection{Two-line shift for massive particles}
      Unfortunately, we can't construct consistent massive BCFW shift for two massive lines as simply as what we do in last subsection. We don't have a natural choice of massless spinors to contract with Weyl-spinor index, since two external lines are both massive particles. It means that the general form of two-line shift doesn't exist. We must choose a specific spinor or reference frame to write down a particular shift. For example, we can only shift one of the helicity states instead of both \cite{Herderschee:2019dmc,Franken:2019wqr} in a special frame.

      Now we still introduce two unknowns $\eta^I$ and $\zeta^J$ in $[\mathbf{i},\mathbf{j}\rangle$-shift, so the shifted spinors are
      \begin{equation}
        |\mathbf{\hat{i}}^I]=|\mathbf{i}^I]+z\eta^I\zeta^J|\mathbf{j}_J],\quad
        |\mathbf{\hat{j}}^J\rangle=|\mathbf{j}^J\rangle+z\eta^I\zeta^J|\mathbf{i}_I\rangle.
      \end{equation}

      The shift vectors are $2r_i^\mu=-2r_j^\mu=\eta^I\zeta^J\langle\mathbf{i}_I|\gamma^\mu|\mathbf{j}_J]$, so condition \eqref{con3} leads to
      \begin{equation}
        \begin{aligned}
          2p_i\cdot r_i &=\eta^I\zeta^J\langle\mathbf{i}_I|p_i|\mathbf{j}_J]= m_i\eta^I\zeta^J[\mathbf{i}_I\mathbf{j}_J]=0,\\
          2p_j\cdot r_j &=-\eta^I\zeta^J\langle\mathbf{i}_I|p_j|\mathbf{j}_J]=m_j\eta^I\zeta^J\langle \mathbf{i}_I\mathbf{j}_J\rangle=0.
        \end{aligned}
      \end{equation}
      The tensor $\eta^I\zeta^J$ has three degrees of freedom, since its determinant is zero. Two equations are not enough to determine all degrees of freedom, so we should choose $\eta^I$ and $\eta^J$ from other information. The massive momenta of two particles can be written as different linear combinations of two null vectors $l_i$ and $l_j$:
      \begin{equation}
        \begin{aligned}
          p_i &=l_i+\alpha_j l_j,\\
          p_j &=l_j+\alpha_i l_i,
        \end{aligned}
      \end{equation}
      where $\alpha_i$ and $\alpha_j$ are coefficient. We choose $\eta^I=[l_j\mathbf{i}^I]/[l_jl_i]$, $\zeta^J=\langle \mathbf{j}^Jl_i\rangle/\langle l_jl_i\rangle$. We can verify condition \ref{con3},
      \begin{equation}
        \begin{aligned}
          \eta^I\zeta^J[\mathbf{i}_I\mathbf{j}_J] &=-\frac{[l_j\mathbf{i}^I][\mathbf{i}_I|p_j|l_i\rangle}{[l_jl_i]\langle l_jl_i\rangle}=-\frac{[l_j\mathbf{i}^I][\mathbf{i}_Il_j]}{[l_jl_i]}=0,\\
          \eta^I\zeta^J\langle \mathbf{i}_I\mathbf{j}_J\rangle &=\frac{[l_j|p_i|\mathbf{j}_J\rangle\langle\mathbf{j}^J l_i\rangle}{[l_jl_i]\langle l_jl_i\rangle}=\frac{\langle l_i\mathbf{j}_J \rangle\langle\mathbf{j}^J l_i\rangle}{\langle l_jl_i\rangle}=0.
        \end{aligned}
      \end{equation}
      Now the shift can be written as
      \begin{equation} \begin{aligned}
        |\mathbf{\hat{i}}^I]=|\mathbf{i}^I]-z\frac{[l_j\mathbf{i}^I]}{[l_jl_i]}|l_j],\quad
        |\mathbf{\hat{j}}^J\rangle=|\mathbf{j}^J\rangle+z\frac{\langle \mathbf{j}^Jl_i\rangle}{\langle l_jl_i\rangle}|l_i\rangle.
      \end{aligned} \end{equation}

    \subsubsection{Three-line Risage-type shifts for massive particles}
      We introduce three unknowns $\eta^I$, $\zeta^J$ and $\xi^K$. Since they are all massive, there is a permutation symmetry between these shifted lines. The shifted spinors are
      \begin{equation} \begin{aligned}
        |\mathbf{\hat{i}}^I]&=|\mathbf{i}^I]+z\eta^I\zeta^J\xi^K\langle \mathbf{j}_J\mathbf{k}_K\rangle|X],\\
        |\mathbf{\hat{j}}^J]&=|\mathbf{j}^J]+z\eta^I\zeta^J\xi^K\langle \mathbf{k}_K\mathbf{i}_I\rangle|X],\\
        |\mathbf{\hat{k}}^K]&=|\mathbf{k}^K]+z\eta^I\zeta^J\xi^K\langle \mathbf{i}_I\mathbf{j}_J\rangle|X].\\
      \end{aligned} \end{equation}
      The shift vectors are
      \begin{equation} \begin{aligned}
        2r_i^\mu&=\eta^I\zeta^J\xi^K\langle\mathbf{j}_J\mathbf{k}_K\rangle\langle\mathbf{i}_I|\gamma^\mu|X],\\
        2r_j^\mu&=\eta^I\zeta^J\xi^K\langle\mathbf{k}_K\mathbf{i}_I\rangle\langle\mathbf{j}_J|\gamma^\mu|X],\\
        2r_k^\mu&=\eta^I\zeta^J\xi^K\langle\mathbf{i}_I\mathbf{j}_J\rangle\langle\mathbf{k}_K|\gamma^\mu|X],\\
      \end{aligned} \end{equation}
      so condition \eqref{con3} leads to
      \begin{equation}
        \begin{aligned}
          2p_i\cdot r_i &=\eta^I\zeta^J\xi^K\langle \mathbf{j}_J\mathbf{k}_K\rangle\langle\mathbf{i}_I|p_i|X]= m_i\eta^I\zeta^J\xi^K\langle \mathbf{j}_J\mathbf{k}_K\rangle[\mathbf{i}_IX]=0,\\
          2p_j\cdot r_j &=\eta^I\zeta^J\xi^K\langle \mathbf{k}_K\mathbf{i}_I\rangle\langle\mathbf{j}_J|p_j|X]= m_j\eta^I\zeta^J\xi^K\langle \mathbf{k}_K\mathbf{i}_I\rangle[\mathbf{j}_JX]=0,\\
          2p_k\cdot r_k &=\eta^I\zeta^J\xi^K\langle \mathbf{i}_I\mathbf{j}_J\rangle\langle\mathbf{k}_K|p_k|X]= m_k\eta^I\zeta^J\xi^K\langle \mathbf{i}_I\mathbf{j}_J\rangle[\mathbf{k}_KX]=0.\\
        \end{aligned}
      \end{equation}

      The solutions are simple, $\eta^I=[\mathbf{i}^IX]$, $\zeta^J=[\mathbf{j}^JX]$, $\xi^K=[\mathbf{k}^KX]$.

    \subsubsection{Three-line BCFW-type shifts for massive particles}
      We introduce three unknowns $\eta^I$, $\zeta^J$ and $\xi^K$, so the shifted spinors are
      \begin{equation} \begin{aligned}
        |\mathbf{\hat{i}}^I]&=|\mathbf{i}^I]+z\eta^I\zeta^J\xi^K\langle\mathbf{j}_JX\rangle |\mathbf{k}_K],\\
        |\mathbf{\hat{j}}^J]&=|\mathbf{j}^J]+z\eta^I\zeta^J\xi^K\langle X\mathbf{i}_I\rangle|\mathbf{k}_K],\\
        |\mathbf{\hat{k}}^K\rangle&=|\mathbf{k}^K\rangle-z\eta^I\zeta^J\xi^K\langle \mathbf{i}_I\mathbf{j}_J \rangle|X\rangle.
      \end{aligned} \end{equation}
      The shift vectors are
      \begin{equation} \begin{aligned}
        2r_i^\mu&=\eta^I\zeta^J\xi^K\langle \mathbf{j}_JX\rangle\langle\mathbf{i}_I|\gamma^\mu|\mathbf{k}_K],\\ 2r_j^\mu&=\eta^I\zeta^J\xi^K\langle X\mathbf{i}_I\rangle\langle\mathbf{j}_J|\gamma^\mu|\mathbf{k}_K],\\ 2r_k^\mu&=\eta^I\zeta^J\xi^K\langle \mathbf{i}_I\mathbf{j}_J\rangle\langle X|\gamma^\mu|\mathbf{k}_K],
      \end{aligned} \end{equation}
      so condition \eqref{con3} leads to
      \begin{equation}
        \begin{aligned}
          2p_i\cdot r_i &=\eta^I\zeta^J\xi^K\langle \mathbf{j}_JX\rangle\langle\mathbf{i}_I|p_i|\mathbf{k}_K]= m_i\eta^I\zeta^J\xi^K\langle \mathbf{j}_JX\rangle[\mathbf{i}_I\mathbf{k}_K]=0,\\
          2p_j\cdot r_j &=\eta^I\zeta^J\xi^K\langle X\mathbf{i}_I\rangle\langle\mathbf{j}_J|p_j|\mathbf{k}_K]= m_j\eta^I\zeta^J\xi^K\langle X\mathbf{i}_I\rangle[\mathbf{j}_J\mathbf{k}_K]=0,\\
          2p_k\cdot r_k &=\eta^I\zeta^J\xi^K\langle \mathbf{i}_I\mathbf{j}_J\rangle\langle X|p_k|\mathbf{k}_K]= -m_k\eta^I\zeta^J\xi^K\langle \mathbf{i}_I\mathbf{j}_J\rangle\langle X\mathbf{k}_K\rangle=0.\\
        \end{aligned}
      \end{equation}

      We can choose $\eta^I=[\mathbf{i}^I\mathbf{k}^K]\xi_K, \zeta^J=[\mathbf{j}^J\mathbf{k}^K]\xi_K, \xi^K=\langle X\mathbf{k}^K\rangle$. However, it isn't the final solution. After substituting $\xi^K=\langle X\mathbf{k}^K\rangle$, we get the final result: $\eta^I=[\mathbf{i}^I|p_k|X\rangle, \zeta^J=[\mathbf{j}^J|p_k|X\rangle$.

  \subsection{Explicit form of three-line shifts}
    In previous discussions, we figured out solutions for all two and three-line shifts. To simplify the expressions of these shifts, we define some new Weyl spinors,
    \begin{equation}
      \begin{aligned}
        |k_m\rangle &=p_m|k],\\
        |X_m\rangle &=p_m|X],\\
        |X_{m,n}\rangle &=p_m|X_n]=p_mp_n|X\rangle,
      \end{aligned}
    \end{equation}
    where $p_m$ and $p_n$ correspond to massive particles. The simplified expressions of three-line shifts are shown in table \ref{solution}, in which the massive spinor helicity variables are denoted in \textbf{BOLD} notation. Now we can write down massless three-line shifts \eqref{3risager} and \eqref{3bcfw} and use replacements listed in table \ref{solution} to rederive three-line shifts for all masses.

    \begin{table}[ht]
      \begin{center}
        \makebox[\linewidth][c]{
        \begin{tabular}{|c|c|c|c|c|}
          \hline
          External Legs & Type & Shifted Spinors & Shift-Vectors &  Replacement \\ \hline
          \multirow{3}*{\makecell[c]{1 massive\\2 massless}} & Risager &
          \makecell[l]{$|\mathbf{\hat{i}}]=|\mathbf{i}]+z\langle jk\rangle[\mathbf{i}X]|X]$\\$|\hat{j}]=|j]+z\langle kX_i\rangle|X]$\\$|\hat{k}]=|k]+z\langle X_ij\rangle|X]$} &
          \makecell[l]{$r_i^\mu=\langle jk\rangle\langle X_i|\gamma^\mu|X]$\\$r_j^\mu=\langle kX_i\rangle\langle j|\gamma^\mu|X]$\\$r_k^\mu=\langle X_ij\rangle\langle k|\gamma^\mu|X]$} &
          \makecell[l]{$|i\rangle\rightarrow|X_i\rangle$}\\ \cline{2-5}
          & \multirow{2}*{BCFW} &
          \makecell[l]{$|\mathbf{\hat{i}}]=|\mathbf{i}]+z\langle Xj\rangle[\mathbf{i}k]|k]$\\$|\hat{j}]=|j]+z\langle k_iX\rangle|k]$\\$|\hat{k}\rangle=|k\rangle+z\langle jk_i\rangle|X\rangle$} &
          \makecell[l]{$r_i^\mu=\langle Xj\rangle\langle k_i|\gamma^\mu|k]$\\$r_j^\mu=\langle k_iX\rangle\langle j|\gamma^\mu|k]$\\$r_k^\mu=\langle jk_i\rangle\langle X|\gamma^\mu|k]$} &
          \makecell[l]{$|i\rangle\rightarrow|k_i\rangle$}\\ \cline{3-5}
          & &
          \makecell[l]{$|\hat{i}]=|i]+z\langle Xj\rangle|X_k]$\\$|\hat{j}]=|j]+z\langle iX\rangle|X_k]$\\$|\mathbf{\hat{k}}\rangle=|\mathbf{k}\rangle+z\langle ji\rangle\langle \mathbf{k}X\rangle|X\rangle$} &
          \makecell[l]{$r_i^\mu=\langle Xj\rangle\langle i|\gamma^\mu|X_k]$\\$r_j^\mu=\langle iX\rangle\langle j|\gamma^\mu|X_k]$\\$r_k^\mu=\langle ji\rangle\langle X|\gamma^\mu|X_k]$} &
          \makecell[l]{$|k]\rightarrow|X_k]$}\\ \hline
          \multirow{3}*{\makecell[c]{2 massive\\1 massless}} & Risager &
          \makecell[l]{$|\mathbf{\hat{i}}]=|\mathbf{i}]+z\langle X_jk\rangle[\mathbf{i}X]|X]$\\$|\mathbf{\hat{j}}]=|\mathbf{j}]+z\langle kX_i\rangle[\mathbf{j}X]|X]$\\$|\hat{k}]=|k]+z\langle X_iX_j\rangle|X]$} &
          \makecell[l]{$r_i^\mu=\langle X_jk\rangle\langle X_i|\gamma^\mu|X]$\\$r_j^\mu=\langle kX_i\rangle\langle X_j|\gamma^\mu|X]$\\$r_k^\mu=\langle X_iX_j\rangle\langle k|\gamma^\mu|X]$} &
          \makecell[l]{$|i\rangle\rightarrow|X_i\rangle$\\$|j\rangle\rightarrow|X_j\rangle$}\\ \cline{2-5}
          & \multirow{2}*{BCFW} &
          \makecell[l]{$|\mathbf{\hat{i}}]=|\mathbf{i}]+z\langle Xk_j\rangle[\mathbf{i}k]|k]$\\$|\mathbf{\hat{j}}]=|\mathbf{j}]+z\langle k_iX\rangle[\mathbf{j}k]|k]$\\$|\hat{k}\rangle=|k\rangle+z\langle k_jk_i\rangle|X\rangle$} &
          \makecell[l]{$r_i^\mu=\langle Xk_j\rangle\langle k_i|\gamma^\mu|k]$\\$r_j^\mu=\langle k_iX\rangle\langle k_j|\gamma^\mu|k]$\\$r_k^\mu=\langle k_jk_i\rangle\langle X|\gamma^\mu|k]$} &
          \makecell[l]{$|i\rangle\rightarrow|k_i\rangle$\\$|j\rangle\rightarrow|k_j\rangle$}\\ \cline{3-5}
          & &
          \makecell[l]{$|\mathbf{\hat{i}}]=|\mathbf{i}]+z\langle Xj\rangle[\mathbf{i}X_k]|X_k]$\\$|\hat{j}]=|j]+z\langle X_{k,i}X\rangle|X_k]$\\$|\mathbf{\hat{k}}\rangle=|\mathbf{k}\rangle+z\langle jX_{k,i}\rangle\langle\mathbf{k}X\rangle|X\rangle$} &
          \makecell[l]{$r_i^\mu=\langle Xj\rangle\langle X_{k,i}|\gamma^\mu|X_k]$\\$r_j^\mu=\langle X_{k,i}X\rangle\langle j|\gamma^\mu|X_k]$\\$r_k^\mu=\langle jX_{k,i}\rangle\langle X|\gamma^\mu|X_k]$} &
          \makecell[l]{$|i\rangle\rightarrow|X_{k,i}\rangle$\\$|k]\rightarrow|X_k]$}\\ \hline
          \multirow{2}*{3 massive} & Risager &
          \makecell[l]{$|\mathbf{\hat{i}}]=|\mathbf{i}]+z\langle X_jX_k\rangle[\mathbf{i}X]|X]$\\$|\mathbf{\hat{j}}]=|\mathbf{j}]+z\langle X_kX_i\rangle[\mathbf{j}X]|X]$\\$|\mathbf{\hat{k}}]=|\mathbf{k}]+z\langle X_iX_j\rangle[\mathbf{k}X]|X]$} &
          \makecell[l]{$r_i^\mu=\langle X_jX_k\rangle\langle X_i|\gamma^\mu|X]$\\$r_j^\mu=\langle X_kX_i\rangle\langle X_j|\gamma^\mu|X]$\\$r_k^\mu=\langle X_iX_j\rangle\langle X_k|\gamma^\mu|X]$} &
          \makecell[l]{$|i\rangle\rightarrow|X_i\rangle$\\$|j\rangle\rightarrow|X_j\rangle$\\$|k\rangle\rightarrow|X_k\rangle$}\\ \cline{2-5}
          & BCFW &
          \makecell[l]{$|\mathbf{\hat{i}}]=|\mathbf{i}]+z\langle XX_{k,j}\rangle[\mathbf{i}X_k]|X_k]$\\$|\mathbf{\hat{j}}]=|\mathbf{j}]+z\langle X_{k,i}X\rangle[\mathbf{j}X_k]|X_k]$\\$|\mathbf{\hat{k}}\rangle=|\mathbf{k}\rangle+z\langle X_{k,j}X_{k,i}\rangle\langle \mathbf{k}X\rangle|X\rangle$} &
          \makecell[l]{$r_i^\mu=\langle XX_{k,j}\rangle\langle X_{k,i}|\gamma^\mu|X_k]$\\$r_j^\mu=\langle X_{k,i}X\rangle\langle X_{k,j}|\gamma^\mu|X_k]$\\$r_k^\mu=\langle X_{k,j}X_{k,i}\rangle\langle X|\gamma^\mu|X_k]$} &
          \makecell[l]{$|i\rangle\rightarrow|X_{k,i}\rangle$\\$|j\rangle\rightarrow|X_{k,j}\rangle$\\$|k]\rightarrow|X_k]$}\\ \hline
        \end{tabular}}
      \end{center}
      \caption{Three-line shifts for all masses, where the little-group indices are suppressed.}
      \label{solution}
    \end{table}

    Furthermore, the notation $|X_{m,n}\rangle$ is not necessary in the expressions. We can rewrite $|X_{m,n}\rangle$ and $|X\rangle$ in terms of $|Y]=|X_n]$,
    \begin{equation}
      \begin{aligned}
        |X_{m,n}\rangle &=p_m|X_n]=p_m|Y\rangle=|Y_m\rangle,\\
        |X\rangle &= \frac{p_np_n|X\rangle}{m^2_n}=\frac{p_n|Y]}{m^2_n}=\frac{|Y_n\rangle}{m^2_n}.
      \end{aligned}
    \end{equation}

    For example, we set $|Y]=|X_k]$. The all-massive BCFW-type three-line shifts reduce to
    \begin{equation}
      \begin{aligned}
        |\mathbf{\hat{i}}] &=|\mathbf{i}]+z\langle Y_kY_j\rangle[\mathbf{i}Y]|Y],\\
        |\mathbf{\hat{j}}] &=|\mathbf{j}]+z\langle Y_iY_k\rangle[\mathbf{j}Y]|Y],\\
        |\mathbf{\hat{k}}\rangle &=|\mathbf{k}\rangle+z\langle Y_jY_i\rangle\langle \mathbf{k}X\rangle|Y_k\rangle.
      \end{aligned}
    \end{equation}
    This expression coincides with \cite{Franken:2019wqr}.

\section{Feynman rules in the large-\texorpdfstring{$z$}{z} limit}
  \label{sec3}
  There is no general constructive method to give an expression of the contribution $B_n$ in eq.\,\eqref{rere}, so the recursion relations hold as long as $A(z\rightarrow\infty)=0$. To investigate the validity of recursion relations, there are many works on the large-$z$ behavior of tree-level massless amplitudes in various shifts \cite{ArkaniHamed:2008yf,Cheung:2008dn,Cohen:2010mi,Cheung:2015cba}. In refs. \cite{ArkaniHamed:2008yf,Cheung:2008dn} the authors focused on the BCFW recursion relations and used background field method to show that, in a theory of spin $\le1$, any massless amplitudes with at least one gluon is constructible. We want to examine whether this argument is applicable to the massive case, so the steps in their proof should be carefully reconsidered.

  We label two external particles of massless amplitudes $A_n$ by 1 and 2. Their momenta are chosen to be deformed,
  \begin{equation}
    \hat{p}_1=p_1+zr,\quad \hat{p}_2=p_2-zr,
  \end{equation}
  which corresponds to eq.\,\eqref{2bcfw}. In the background field method, the large-$z$ behavior of amplitudes $A_n(z)$ have a nice physical interpretation. We take particles 1 and 2 to be incoming and outgoing, so this process can be interpreted as a hard particle shooting through a soft background. In the hard limit $z\rightarrow\infty$, the $z$-independent soft physics is treated as a classical background, while the large-$z$ behavior of amplitudes is completely determined by the hard fluctuations.

  Now, let's discuss the $z$-dependent propagators, vertices and external legs separately. We will see differences in the case when hard fluctuations correspond to massive particles.

  \subsection{Hard propagators}
    The first problem is how massive propagators scale at large-$z$. Both massive fermions and scalar propagators scale as the same as massless propagators, while a massive vector propagator goes as $\mathcal{O}(z)$ :
    \begin{equation}
      \Pi^{\mu\nu}
      =\frac{g^{\mu\nu}-\frac{\hat{p}^\mu\hat{p}^\nu}{m^2}}{\hat{p}^2-m^2} =\frac{g^{\mu\nu}-\frac{(p^\mu+zr^\mu)(p^\nu+zr^\nu)}{m^2}}{p^2+2zp\cdot r-m^2} \overset{z\rightarrow\infty}{=}-z\frac{r^\mu r^\nu}{2m^2p\cdot r}.
      \label{hardp}
    \end{equation}

    Notice that a massless vector propagator goes as $\mathcal{O}(1/z)$, so we should distinguish massless and massive vectors in the following discussion. As in table \ref{propa}, we use single and double wavy lines to make a distinction between massless and massive vectors propagators.

    \begin{table}[ht]
      \begin{center}
        \begin{tabular}{|c|c|c|}
          \hline
          particle & massless & massive \\ \hline
          scalar & \multicolumn{2}{|c|}{
            \begin{fmfgraph*}(50,10)
              \fmfleft{i1} \fmfright{o1}
              \fmf{dashes}{i1,o1}
            \end{fmfgraph*}
          }\\ \hline
          fermion & \multicolumn{2}{|c|}{
            \begin{fmfgraph*}(50,10)
              \fmfleft{i1} \fmfright{o1}
              \fmf{plain}{i1,o1}
            \end{fmfgraph*}
          }\\ \hline
          vector boson &
          \begin{fmfgraph*}(50,10)
            \fmfleft{i1} \fmfright{o1}
            \fmf{wiggly}{i1,o1}
          \end{fmfgraph*} &
          \begin{fmfgraph*}(50,10)
            \fmfleft{i1} \fmfright{o1}
            \fmf{dbl_wiggly}{i1,o1}
          \end{fmfgraph*} \\ \hline
        \end{tabular}
      \end{center}
      \caption{Propagators}
      \label{propa}
    \end{table}

    Here we introduce a diagrammatic expression to represent the numerator in propagator \eqref{hardp}:
    \begin{equation}
      g^{\mu\nu}-\frac{\hat{p}^\mu\hat{p}^\nu}{m^2}\equiv
      \parbox[c]{1.1cm}{\Hardline{\mu}{\nu}}+\frac{1}{m^2}
      \parbox[c]{1.1cm}{\Hardline{\mu}{\hat{p}}}
      \parbox[c]{1.1cm}{\Hardline{\hat{p}}{\nu}},
    \end{equation}
    where the line segment and the letters attached with it, which are either indices or momentum $\hat{p}$, compose a representation of Lorentz group. The first diagram represents a symmetric tensor $g^{\mu\nu}$, in which the line segment connects two Lorentz indices. The second diagram represents two Lorentz vectors, where the line segments connect the shifted momentum $\hat{p}$ to Lorentz indices.

    Furthermore, Einstein summation also has a diagrammatic representation,
    \begin{equation}
      \sum_{\mu}
      \parbox[c]{1.1cm}{\Hardline{p}{\mu}}
      \parbox[c]{1.1cm}{\Hardline{\mu}{p}}=
      \parbox[c]{1.1cm}{\Hardline{p}{p}}=p^2=m^2,
    \end{equation}
    where two same indices are equivalent to a line segment. Therefore, the first diagram reduces to a line segment with two momenta $p$ attached to it, which represents Lorentz scalar $p^2$.

  \subsection{External polarizations}
    External spinors are constructed in the little-group notation,
    \begin{equation}
      \bar{u}^I(p)=
      \begin{pmatrix}
        [\mathbf{p}^I| &
        \langle\mathbf{p}^I|
      \end{pmatrix},\quad
      v^I(p)=
      \begin{pmatrix}
        |\mathbf{p}^I] \\
        |\mathbf{p}^I\rangle
      \end{pmatrix},
    \end{equation}
    where the little-group indices $I=1,2$ characterize two different solutions of the Dirac equation.

    For massive vector bosons, the polarization vectors transform under the three-dimensional tensor representation of the little group,
    \begin{equation}
      \epsilon_\mu^{I_1I_2}(p)=-\frac{1}{\sqrt{2}m}\left[\mathbf{p}^{I_1}|\gamma_\mu|\mathbf{p}^{I_2}\right\rangle.
      \label{massivepolar}
    \end{equation}
    However, this little-group covariant expression is not convenient when we are talking about amplitudes in the large-$z$ limit. Usually we don't choose the rest frame of massive particles as a reference system, since any shift should include more than one external leg. It implies that we can choose the spin axis along the 3-momentum direction and write down the spin state for massive vector particles in terms of $\epsilon^{I_1I_2}$:
    \begin{equation}
      \epsilon_i^{+}=\epsilon^{11}(p_i),\quad
      \epsilon_i^{0}=\frac{1}{2}\left(\epsilon^{12}(p_i)+\epsilon^{21}(p_i)\right),\quad
      \epsilon_i^{-}=-\epsilon^{22}(p_i),
    \end{equation}
    where we ignore Lorentz indices.

    \subsubsection{\texorpdfstring{$[1,\mathbf{2}\rangle$}{[1,2>}-shift}
      To consider two-line shifts, we take particles 1 and 2 to be massless and massive respectively. Particle 1 is always a gauge boson, while particle 2 can be any massive particle whose spin $\le1$. Since the large-$z$ behavior is independent of the reference system, we can choose the center-of-mass frame of the particles for simplicity. The momenta of particle 1 and 2 and shift vectors become
      \begin{equation}
        \begin{aligned}
          p_1^\mu=(1,0,0,1),\quad p_2^\mu=(\sqrt{m^2+1},0,0,-1),\quad r^\mu=(0,-1,-i,0),
        \end{aligned}
        \label{p1p2r}
      \end{equation}
      where the 3-momenta are normalized. The shift vector $r$ is basically the same as the polarization vectors for the real momentum $p_1$.

      In pure Yang-Mills theory, the shift vector $r$ is enough to construct the polarization vectors. We add a new vector $\bar{p}_2$ to give the longitudinal polarization, whose spatial components point opposite to the direction of spatial components of $p_2$. Similarly, we construct a new null vector $\bar{p}_1$. The expressions of $\bar{p}_1$ and $\bar{p}_2$ are,
      \begin{equation}
        \bar{p}_1^\mu=(1,0,0,-1),\quad \bar{p}_2^\mu=(1,0,0,-\sqrt{m^2+1}).
      \end{equation}

      We choose $[1,\mathbf{2}\rangle$-shift \eqref{ij2}. The shifted polarization vectors of particle 1 are
      \begin{equation}
        \begin{aligned}
          \hat{\epsilon}_1^+=r^*+z\bar{p}_1,\quad \hat{\epsilon}_1^-=r,
        \end{aligned}
        \label{polar1}
      \end{equation}
      which are the same as in the all-massless case. They should stay orthogonal to momentum $\hat{p}_1$ and their product $(\epsilon_1^+\epsilon_1^-)$ is maintained. The Ward identity is still valid for complexified amplitudes:
      \begin{equation}
        \hat{p}_{1}^\mu\hat{A}_\mu(z)=(p_1^\mu+zr^\mu)\hat{A}_\mu(z)=0.
      \end{equation}
      Therefore, we can use it to replace the negative polarization,
      \begin{equation}
        \hat{\epsilon}_1^-\rightarrow -\frac{1}{z}p_1.
        \label{replacep1}
      \end{equation}

      If particle 2 is a vector boson, we choose the $z$ axis as the spin direction, and the shifted external polarizations are
      \begin{equation}
        \hat{\epsilon}_2^+=r,\quad
        \hat{\epsilon}_2^-=r^*-zCp_1,\quad
        \hat{\epsilon}_2^0=\frac{\bar{p}_2}{m}-z\frac{r}{m},\\
        \label{polar2}
      \end{equation}
      where $C=2/(\sqrt{m^2+1}+1)$. Since we focus on the large-$z$ behavior of amplitudes, the overall factor $1/m$ in the expression of $\hat{\epsilon}_2^0$ can be ignored. The polarizations are modified appropriately to remain normalized to unity and orthogonal to $\hat{p}_2$. What's more, the longitudinal polarization should be orthogonal to other two transverse polarizations. A more detailed discussion is performed in appendix \ref{AppB}. Using Goldstone boson equivalence theorem, the scaling behaviour of massive vector bosons can also be improved \cite{Franken:2019wqr}. However, this improvement changes type of particles, so we don't apply it in the following analysis.

      If particle 2 is a fermion, we need the shifted Dirac spinors,
      \begin{equation}
        \hat{\bar{u}}^J_2=\bar{u}^J_2-z\bar{u}^-_1 \langle 1\mathbf{2}^J\rangle,\quad
        \hat{v}^J_2=v^J_2-z v^-_1 \langle 1\mathbf{2}^J\rangle.\\
        \label{shiftedf}
      \end{equation}
      Although particle 1 is a gauge boson, we still use $\bar{u}^-_1=\langle1|$ and $v^-_1=|1\rangle$ for consistency.

      \subsubsection{\texorpdfstring{$[\mathbf{1},\mathbf{2}\rangle$}{[1,2>]}-shift}
      Particles 1 and 2 have equal mass. The momenta of them become
      \begin{equation}
        \begin{aligned}
          p_1^\mu=(\sqrt{m^2+1},0,0,1),\quad p_2^\mu=(\sqrt{m^2+1},0,0,-1).
        \end{aligned}
      \end{equation}
      We set the shift vector $r$ to be the same as in eq.\,\eqref{p1p2r}. Now both $p_1$ and $p_2$ are time-like vectors, we need two vectors $\bar{p}_1$ and $\bar{p}_2$ to give the longitudinal polarizations,
      \begin{equation}
        \begin{aligned}
          \bar{p}_1^\mu=(1,0,0,\sqrt{m^2+1}),\quad \bar{p}_2^\mu=(1,0,0,-\sqrt{m^2+1}).
        \end{aligned}
      \end{equation}
      Here we only discuss the case that both particles 1 and 2 are massive vector bosons. The shifted external legs of massive vectors are
      \begin{equation}
        \begin{aligned}
          \hat{\epsilon}_1^+=r^*+zC^2\frac{p_2+\bar{p}_2}{2},\quad
          \hat{\epsilon}_1^-=r,\quad
          \hat{\epsilon}_1^0=\frac{\bar{p}_1}{m}+z\frac{r}{m},\\
          \hat{\epsilon}_2^+=r,\quad
          \hat{\epsilon}_2^-=r^*-zC^2\frac{p_1+\bar{p}_1}{2},\quad
          \hat{\epsilon}_2^0=\frac{\bar{p}_2}{m}-z\frac{r}{m},\\
        \end{aligned}
        \label{polar3}
      \end{equation}
      where $C=2/(\sqrt{m^2+1}+1)$. In the high energy limit, $C\rightarrow 1$, $\bar{p}_1\rightarrow p_1$ and $\bar{p}_2\rightarrow p_2$. Therefore, eqs. \eqref{polar1} and \eqref{polar2} become the high energy limit of eq.\,\eqref{polar3}.

  \subsection{\texorpdfstring{$\mathcal{O}(z)$}{O(z)} vertices}
    \label{vertex}
    Along the hard fluctuation, the spin of the hard particle may be changed by $z$-dependent vertices which involve soft background fields. Since we are discussing renormalizable field theory, the $z$-dependent vertices from derivative interactions must be linear in $z$. In massless gauge theory, they are eliminated by choosing appropriate light-cone and $R_\xi$ gauges \cite{ArkaniHamed:2008yf,Elvang:2013cua}. However, the massive vector bosons don't have such degrees of freedom to eliminate these $z$-dependence. In renormalizable field theory, there are two classes of $\mathcal{O}(z)$ vertices that cannot be eliminated by gauge fixing in the massive case: triple vector coupling ($VVV$) and Vector-Vector-Scalar ($VSS$) interaction. For simplicity, we only consider massive vector bosons that have equal mass.

    Since all on-shell 3-point amplitudes in the Standard Model have been figured out\cite{Christensen:2018zcq}, we can translate them into the Feynman rules to find out the vertices. Then we deform these vertices to give their $z$-dependence explicitly.

    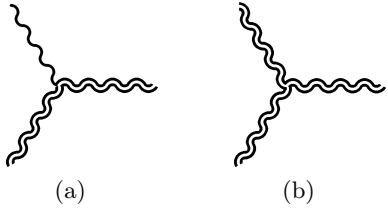
\begin{figure}[htbp]
    	\centering
      \subfloat[]{\label{1A2F}
        \begin{fmfgraph*}(60,60)
          \fmfleft{i1,i2} \fmfright{o1}
          \fmf{dbl_wiggly}{i1,v1}
          \fmf{photon}{i2,v1}
          \fmf{dbl_wiggly}{v1,o1}
        \end{fmfgraph*}}\qquad
      \subfloat[]{
        \begin{fmfgraph*}(60,60)
          \fmfleft{i1,i2} \fmfright{o1}
          \fmf{dbl_wiggly}{i1,v1}
          \fmf{dbl_wiggly}{i2,v1}
          \fmf{dbl_wiggly}{v1,o1}
        \end{fmfgraph*}}\\
    	\caption{$VVV$ vertices with massive vectors}
      \label{VVV}
    \end{figure}

    There are two kinds of possible $VVV$ amplitudes (see figure \ref{VVV}). Since $VVV$ amplitude has three vector external legs, the vertex should be a order-3 Lorentz tensor $V_{k,p,q}^{\mu\nu\lambda}$, which refers to $V$ attached with three line segments in the diagrammatic expression. One kind of vertex includes two massive vectors and one massless vector. Its diagrammatic representations are
    \begin{equation}
      \begin{aligned}
        \parbox[b]{2.8cm}{
        \begin{tikzpicture}
          \draw[thick]
              (0,0) node[circle,inner sep=0pt,draw=white,fill=white]{$\epsilon_i^{I_1I_2}$}
            --(0.9,0) node[circle,inner sep=0pt,draw=white,fill=white]{$V$}
            --(1.85,0) node[circle,inner sep=0pt,draw=white,fill=white]{$\epsilon_j^{J_1J_1}$};
          \draw[thick]
              (0.9,0) node[circle,inner sep=0pt,draw=white,fill=white]{$V$}
            --(0.9,0.7) node[circle,inner sep=0pt,draw=white,fill=white]{$\epsilon_k^+$};
        \end{tikzpicture}\vspace{-0.4cm}}
        =\frac{\sqrt{2}}{m}x\langle\mathbf{i}\mathbf{j}\rangle^2,\qquad
        \parbox[b]{2.8cm}{
        \begin{tikzpicture}
          \draw[thick]
              (0,0) node[circle,inner sep=0pt,draw=white,fill=white]{$\epsilon_i^{I_1I_2}$}
            --(0.9,0) node[circle,inner sep=0pt,draw=white,fill=white]{$V$}
            --(1.85,0) node[circle,inner sep=0pt,draw=white,fill=white]{$\epsilon_j^{J_1J_1}$};
          \draw[thick]
              (0.9,0) node[circle,inner sep=0pt,draw=white,fill=white]{$V$}
            --(0.9,0.7) node[circle,inner sep=0pt,draw=white,fill=white]{$\epsilon_k^-$};
        \end{tikzpicture}\vspace{-0.4cm}}
        =\frac{\sqrt{2}}{m}\frac{1}{x}[\mathbf{i}\mathbf{j}]^2,
      \end{aligned}
      \label{WWA}
    \end{equation}
    where the $x$ factor is introduced by \cite{Arkani-Hamed:2017jhn}, which carries $+1$ helicity. The vertex $\gamma W^+W^-$ in the Standard Model belongs to this kind of vertex.

    Another kind of amplitude includes three massive vectors. Its diagrammatic representation is
    \begin{equation}
      \begin{aligned}
        \parbox[b]{2.8cm}{
        \begin{tikzpicture}
          \draw[thick]
              (0,0) node[circle,inner sep=0pt,draw=white,fill=white]{$\epsilon_i^{I_1I_2}$}
            --(0.9,0) node[circle,inner sep=0pt,draw=white,fill=white]{$V$}
            --(1.85,0) node[circle,inner sep=0pt,draw=white,fill=white]{$\epsilon_j^{J_1J_1}$};
          \draw[thick]
              (0.9,0) node[circle,inner sep=0pt,draw=white,fill=white]{$V$}
            --(0.9,0.8);
          \draw (1.2,0.8) node[circle,inner sep=-1.5pt,draw=white,fill=white]{$\epsilon_k^{K_1K_1}$};
        \end{tikzpicture}\vspace{-0.4cm}}
        &=\frac{[\mathbf{i}\mathbf{j}]\langle\mathbf{j}\mathbf{k}\rangle[\mathbf{k}\mathbf{i}]+\langle\mathbf{i}\mathbf{j}\rangle[\mathbf{j}\mathbf{k}]\langle\mathbf{k}\mathbf{i}\rangle}{\sqrt{2}m^2}
        +\frac{[\mathbf{i}\mathbf{j}][\mathbf{j}\mathbf{k}]\langle\mathbf{k}\mathbf{i}\rangle+\langle\mathbf{i}\mathbf{j}\rangle\langle\mathbf{j}\mathbf{k}\rangle[\mathbf{k}\mathbf{i}]}{\sqrt{2}m^2}\\
        &\;\quad+\frac{\langle\mathbf{i}\mathbf{j}\rangle[\mathbf{j}\mathbf{k}][\mathbf{k}\mathbf{i}]+[\mathbf{i}\mathbf{j}]\langle\mathbf{j}\mathbf{k}\rangle\langle\mathbf{k}\mathbf{i}\rangle}{\sqrt{2}m^2}.\\
      \end{aligned}
      \label{WWW}
    \end{equation}
    There is no such vertex in the Standard Model, because $W$ and $Z$ bosons have different masses.

    Now we give the expression for the $VVV$ vertex $V_{k,p,q}^{\mu\nu\lambda}$,
    \begin{equation}
      \begin{aligned}
        V_{k,p,q}^{\mu\nu\lambda}\equiv
        \parbox[b]{1.8cm}{\Hards{\mu}{V}{\lambda}{\nu}\vspace{-0.15cm}}
        =g^{\mu\nu}(p-k)^\lambda+g^{\nu\lambda}(q-p)^\mu+g^{\lambda\mu}(k-q)^\nu,
      \end{aligned} \label{Vunhat}
    \end{equation}
    where $k+p+q=0$. It is easy to check this expression by dotting it into vector boson polarizations. This manipulation will give the amplitudes in eqs. \eqref{WWA} and \eqref{WWW} again. We can use shifted momenta $k+zr$ and $q-zr$ instead of $k$ and $q$ to deform this vertex. Diagrammatically, shifted vertex is represented as
    \begin{equation}
      \begin{aligned}
        \parbox[b]{1.8cm}{\Hards{\mu}{\hat{V}}{\lambda}{\nu}\vspace{-0.17cm}}
        =\hat{V}_{k+zr,p,q-zr}^{\mu\nu\lambda}
        =V_{k,p,q}^{\mu\nu\lambda}+zR^{\mu\nu\lambda}=
        \parbox[b]{1.8cm}{\Hards{\mu}{V}{\lambda}{\nu}\vspace{-0.13cm}}
        +z\times
        \parbox[b]{1.8cm}{\Hards{\mu}{R}{\lambda}{\nu}\vspace{-0.13cm}},
      \end{aligned}
      \label{Vhat}
    \end{equation}
    where $R^{\mu\nu\lambda}=(-g^{\mu\nu}r^\lambda-g^{\nu\lambda}r^\mu+2g^{\lambda\mu}r^\nu)$. In the diagram, the three lines correspond to three Lorentz indices of the vertex. Two horizontal lines represent hard fluctuations, so the momenta they carry should be shifted (e.g. $k+zr$ and $q-zr$ in eq.\,\eqref{Vhat}).

    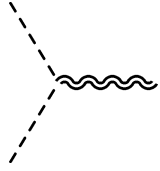
\begin{figure}[htbp]
    	\centering
        \begin{fmfgraph*}(60,60)
          \fmfleft{i1,i2} \fmfright{o1}
          \fmf{dashes}{i1,v1}
          \fmf{dashes}{i2,v1}
          \fmf{dbl_wiggly}{v1,o1}
        \end{fmfgraph*}
    	\caption{$VSS$ vertex with massive vector}
      \label{2s1F}
    \end{figure}

    Next, $VSS$ amplitude (see figure \ref{2s1F}) will give a simpler vertex. Since the only one external vector boson contributes one Lorentz index, the vertex must be a Lorentz vector. Suppose the momenta of scalar bosons are $p$ and $q$, the vertex $V_{p,q}^\mu$ will be
    \begin{equation}
      \parbox[c]{1.2cm}{
      \begin{tikzpicture}
        \draw[thick]
            (0,0)node[circle,inner sep=0pt,draw=white,fill=white] {$V$}
          --(0.8,0)node[circle,inner sep=0pt,draw=white,fill=white] {$\mu$};
      \end{tikzpicture}}=(p-q)^{\mu}.
    \end{equation}
    This vertex can be realized in various BSM models, such as 2HDM, MSSM and the simplest Little Higgs model \cite{Gunion:1989we,He:2017jjx}.  In the last case, there is a Higgs-Goldstone mixing term in the non-linear Lagrangian.

    Notice that the $VSS$ vertex seems to be a substructure of $VVV$ vertex. Diagrammatically, this means
    \begin{equation}
      \parbox[b]{1.8cm}{\Hards{\mu}{V}{\lambda}{\nu}\vspace{-0.13cm}}=
      \parbox[b]{1.8cm}{\begin{tikzpicture}
      \draw[thick]
          (0,0)node[circle,inner sep=0pt,draw=white,fill=white]{$\mu$}
          --(1.4,0)node[circle,inner sep=0pt,draw=white,fill=white]{$\lambda$};
      \draw[thick]
          (0.7,0.25)node[circle,inner sep=0pt,draw=white,fill=white]{$V$}
          --(0.7,0.75)node[circle,inner sep=0pt,draw=white,fill=white]{$\nu$};
      \end{tikzpicture}\vspace{-0.15cm}}+
      \parbox[b]{1.8cm}{\begin{tikzpicture}
      \draw[thick]
          (0,0)node[circle,inner sep=0pt,draw=white,fill=white]{$\mu$}
          --(0.6,0)
          --(0.6,0.75)node[circle,inner sep=0pt,draw=white,fill=white]{$\nu$};
      \draw[thick]
          (0.85,0)node[circle,inner sep=0pt,draw=white,fill=white]{$V$}
          --(1.4,0)node[circle,inner sep=0pt,draw=white,fill=white]{$\lambda$};
      \end{tikzpicture}\vspace{-0.15cm}}+
      \parbox[b]{1.8cm}{\begin{tikzpicture}
      \draw[thick]
          (1.4,0)node[circle,inner sep=0pt,draw=white,fill=white]{$\lambda$}
          --(0.8,0)
          --(0.8,0.75)node[circle,inner sep=0pt,draw=white,fill=white]{$\nu$};
      \draw[thick]
          (0,0)node[circle,inner sep=0pt,draw=white,fill=white]{$\mu$}
          --(0.55,0)node[circle,inner sep=0pt,draw=white,fill=white]{$V$};
      \end{tikzpicture}\vspace{-0.15cm}}.
      \label{substructure}
    \end{equation}
    Therefore, the large-$z$ behavior of $VSS$ vertex is contained in $VVV$ vertex.

\section{Large-\texorpdfstring{$z$}{z} behavior of \texorpdfstring{$n$}{n}-point amplitudes}
\label{sec4}
For massless amplitudes, appropriate gauges eliminate all large-$z$ contributions from derivative interactions except in the so-called ``unique diagrams". Therefore, the large-$z$ behavior of amplitudes depends on the number of hard propagators. If there are more hard boson propagators, the amplitudes will be suppressed by higher powers of $z$. As for hard fermion propagators, the contribution can be simplified by using anti-commuted gamma matrices and $\slashed{r}\slashed{r}=r^2=0$. At last, the hard polarizations are dotted into the sum of all contributions.

In the massive case, we have seen that the $\mathcal{O}(z)$ vertices are not eliminated completely, so the diagrams with hard vector propagators cannot be ignored. Thus, the large-$z$ behavior in the massive case should be evaluated directly instead of estimated.

One may wonder if the Goldstone boson equivalence theorem \cite{Lee:1977eg,Chanowitz:1985hj} is useful in the complex deformation. In the high energy limit, this theorem treats the longitudinal modes of vector bosons as Goldstone bosons to simplify calculations. However, the large-$z$ limit is not exactly the same as the high energy limit. Actually, it has been used in amplitudes under the all-line shift \cite{Cohen:2010mi}, but there are also many cases where it doesn't work. Consider the following amplitude in the large-$z$ limit with particles 1 and 2 shifted:
\begin{equation}
  \begin{gathered}
    \begin{fmfgraph*}(90,50)
      \fmfbottom{i1,o1}\fmfv{l=1,l.a=70,l.d=.06w}{i1}
      \fmf{photon}{i1,v1}
      \fmf{dbl_wiggly}{v1,o1}\fmfv{l=2,l.a=110,l.d=.05w}{o1}
      \fmffreeze
      \fmftop{i2}\fmfv{l=3,l.a=-30,l.d=.08w}{i2}
      \fmf{dbl_wiggly}{v1,i2}
    \end{fmfgraph*}
  \end{gathered}\quad=\quad
  \begin{gathered}
    \begin{fmfgraph*}(90,50)
      \fmfbottom{i1,o1}
      \fmf{photon}{i1,v1}\fmfv{l=1,l.a=70,l.d=.06w}{i1}
      \fmf{dashes}{v1,o1}\fmfv{l=2,l.a=120,l.d=.06w}{o1}
      \fmffreeze
      \fmftop{i2}\fmfv{l=3,l.a=-30,l.d=.08w}{i2}
      \fmf{dashes}{v1,i2}
    \end{fmfgraph*}
  \end{gathered}
  \times\left(1+\mathcal{O}\left(\frac{m_\Phi^2}{E^2}\right)\right).\\
\end{equation}
where $m_\Phi$ is the mass of scalar bosons and $E$ is the energy of particle 3. Particle 3 is unshifted, so $\frac{m_\Phi^2}{E^2}$ is no longer a negligible quantity. If we insist on applying this expansion, we should sum over the contributions from infinite terms. Thus, we won't use the Goldstone boson equivalence theorem in our analysis.

\subsection{\texorpdfstring{$n$}{n}-point Vector Boson Scattering Amplitudes}
In the massive case, the large-$z$ behavior of amplitudes depends on massive vector bosons, so we can first consider vector boson scattering. Besides particle 1, we set all particles to be massive. The massive vector propagator \eqref{hardp} has two terms. The first term connects vertices while the second term splits the diagram in two. Therefore, the amplitudes can be split into two parts,
\begin{equation} \begin{aligned}
  \hat{A}_n=\hat{A}_n^C+\hat{A}_n^D,
\end{aligned} \end{equation}
where $\hat{A}_n^C$ and $\hat{A}_n^D$ correspond to the connected and disconnected diagrammatic expressions respectively.

When a hard particle shoots through a soft background, it will interact with the classical field more than once. The soft physics is parameterized by currents $B^{\mu_j}$, so the amplitude $\hat{A}_n^C$ becomes
\begin{equation} \begin{aligned}
  \hat{A}_n^C=&N_{\mu_3}B^{\mu_3}_3+\sum_{i=4}^{n}\sum_{\sigma\in S_{i-2}}\frac{N_{\sigma(\mu_3\mu_4\cdots\mu_i)}}{\prod_{j=3}^{i-1} D_{\sigma(j)}}\prod_{j=3}^{i} B^{\mu_j}_i,
  \label{connect1}
\end{aligned} \end{equation}
where the second sum is over all permutations of the labels $(3,4,\dots,i)$. Here $D_j$ is the denominator of shifted propagators. After permutation, it becomes
\begin{equation} \begin{aligned}
  D_{\sigma(j)}&=\left(\hat{p}_1+\sum_{k=3}^j p_{\sigma(k)}\right)^2-m^2=2z\sum_{k=3}^j p_\sigma(k) \cdot r+\mathcal{O}(1).
  \label{capitald}
\end{aligned} \end{equation}

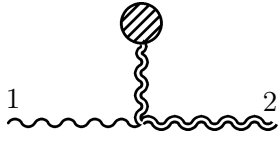
\begin{figure}[htbp]
  \centering
    \subfloat{\label{aVb}
      \begin{fmfgraph*}(100,70)
        \fmfbottom{i1,o1}\fmfv{l=1,l.a=70,l.d=.06w}{i1}
        \fmf{photon}{i1,v1}
        \fmf{dbl_wiggly}{v1,o1}\fmfv{l=2,l.a=110,l.d=.05w}{o1}
        \fmffreeze
        \fmftop{i2}
        \fmf{phantom}{i2,v3,v2}
        \fmf{dbl_wiggly}{v3,v2,v1}
        \fmfblob{.15w}{v3}
      \end{fmfgraph*}}
  \caption{The diagram without hard propagators. The blob represents the soft background.}
  \label{unique}
\end{figure}

The first term in eq.\,\eqref{connect1} corresponds to the diagram without hard propagators (see figure \ref{unique}). Since the hard particle interacts with the classical field only once, the large-$z$ behavior of the numerator should be given by a Lorentz vector $N_{\mu_3}$. Inserting polarizations \eqref{polar2} and \eqref{replacep1}, it becomes
\begin{equation}  \begin{aligned}
  N^{-,+}_{\mu_3}&=
  \parbox[b]{2.1cm}{\Hards{\hat{\epsilon}_1^-}{\hat{V}}{\hat{\epsilon}_2^+}{\mu}\vspace{-0.2cm}}
  =-\parbox[b]{1.8cm}{\Hards{p_1}{R}{r}{\mu}\vspace{-0.15cm}}
  +\mathcal{O}\left(\frac{1}{z}\right)
  \rightarrow\mathcal{O}\left(\frac{1}{z}\right),\\
  N^{-,-}_{\mu_3}&=
  \parbox[b]{2.1cm}{\Hards{\hat{\epsilon}_1^-}{\hat{V}}{\hat{\epsilon}_2^-}{\mu}\vspace{-0.2cm}}\\
  &=zC\parbox[b]{1.9cm}{\Hards{p_1}{R}{p_1}{\mu}\vspace{-0.15cm}}
  -\parbox[b]{0.3cm}{\Bigg(\vspace{-0.2cm}}\parbox[b]{1.9cm}{\Hards{p_1}{R}{r^*}{\mu}\vspace{-0.15cm}}
  -C\parbox[b]{1.9cm}{\Hards{p_1}{V}{p_1}{\mu}\vspace{-0.15cm}}
  \parbox[b]{0.3cm}{\Bigg)\vspace{-0.2cm}}+\mathcal{O}\left(\frac{1}{z}\right)\\
  &=[(r^*r)+C(p_1p_2)]p_1^\mu-\mathcal{O}\left(\frac{1}{z}\right)
  \rightarrow\mathcal{O}\left(\frac{1}{z}\right),\\
  N^{-,0}_{\mu_3}&=
  \parbox[b]{2.1cm}{\Hards{\hat{\epsilon}_1^-}{\hat{V}}{\hat{\epsilon}_2^0}{\mu}\vspace{-0.2cm}}\\
  &=z\parbox[b]{1.9cm}{\Hards{p_1}{R}{r}{\mu}\vspace{-0.15cm}}
  -\parbox[b]{0.3cm}{\Bigg(\vspace{-0.2cm}}\parbox[b]{1.9cm}{\Hards{p_1}{R}{\bar{p}_2}{\mu}\vspace{-0.15cm}}
  -\parbox[b]{1.9cm}{\Hards{p_1}{V}{r}{\mu}\vspace{-0.15cm}}
  \parbox[b]{0.3cm}{\Bigg)\vspace{-0.2cm}}+\mathcal{O}\left(\frac{1}{z}\right)\\
  &=-2[(p_1\bar{p}_2)-(p_1p_2)]r^\mu+\mathcal{O}\left(\frac{1}{z}\right)
  \rightarrow\mathcal{O}\left(\frac{1}{z}\right).
  \label{N3}
\end{aligned} \end{equation}
The leading diagrams are equal to zero. We used $(r^*r)+C(p_1p_2)=0$ and $(p_1\bar{p}_2)=(p_1p_2)$ to cancel the subleading diagrams in $N^{-,+}_{\mu_3}$ and $N^{-,0}_{\mu_3}$. It shows that the first term in eq.\,\eqref{connect1} vanishes in the large-$z$ limit.

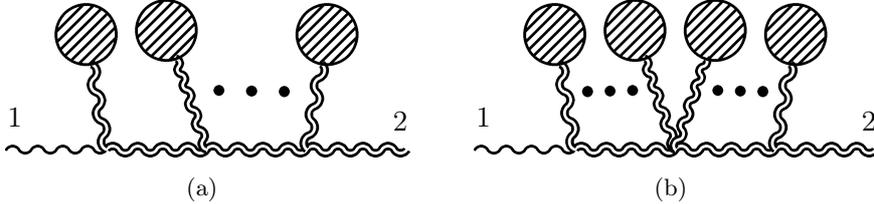
\begin{figure}[htbp]
  \centering
  \subfloat[]{
    \begin{fmfgraph*}(150,50)
      \fmfbottom{b1,b5}\fmfv{l=1,l.a=70,l.d=.06w}{b1}
      \fmf{photon}{b1,b2}
      \fmf{dbl_wiggly}{b2,b3,b4,b5}
      \fmfv{l=2,l.a=110,l.d=.05w}{b5}
      \fmffreeze
      \fmftopn{s}{6}
      \fmf{dbl_wiggly}{s2,b2}
      \fmf{dbl_wiggly}{s3,m1,b3}
      \fmf{dbl_wiggly}{s5,m2,b4}
      \fmfblob{.15w}{s2,s3,s5}
      \fmffreeze
      \fmf{phantom}{m1,m3,m4,m5,m2}
      \fmfv{d.shape=circle,d.f=full,d.size=1.5thick}{m3,m4,m5}
    \end{fmfgraph*}}\qquad
  \subfloat[]{
    \begin{fmfgraph*}(150,50)
      \fmfbottom{b1,b5}\fmfv{l=1,l.a=70,l.d=.06w}{b1}
      \fmf{photon}{b1,b2}
      \fmf{dbl_wiggly}{b2,b3,b4,b5}
      \fmfv{l=2,l.a=110,l.d=.05w}{b5}
      \fmffreeze
      \fmftopn{s}{6}
      \fmf{dbl_wiggly}{s2,m1,b2}
      \fmf{dbl_wiggly}{s3,m2,b3}
      \fmf{dbl_wiggly}{s4,m3,b3}
      \fmf{dbl_wiggly}{s5,m4,b4}
      \fmfblob{.15w}{s2,s3,s4,s5}
      \fmffreeze
      \fmf{phantom}{m1,ma1,ma2,ma3,m2}
      \fmf{phantom}{m3,mb1,mb2,mb3,m4}
      \fmfv{d.shape=circle,d.f=full,d.size=1.5thick}{ma1,ma2,ma3}
      \fmfv{d.shape=circle,d.f=full,d.size=1.5thick}{mb1,mb2,mb3}
    \end{fmfgraph*}}
  \caption{The diagrams that give the leading contributions in the large-$z$ limit. The blobs correspond to soft backgrounds.}
  \label{npt}
\end{figure}

As for the second term in eq.\,\eqref{connect1}, the large-$z$ behavior is given by a Lorentz tensor $N_{\sigma(\mu_3\mu_4\cdots\mu_i)}$. Since this term includes more than three external particles, 4-vertices should be considered. We use a tensor $V_4^{\mu\nu\sigma\rho}$ to represent a 4-vertex. Since there are three ways to contract four polarization vectors, the diagrammatic expression of a 4-vertex should be
\begin{equation}\begin{aligned}
    V_4^{\mu\nu\sigma\rho}\equiv
    \parbox[b]{1.8cm}{
      \begin{tikzpicture}
        \draw[thick]
            (0,0) node[circle,inner sep=0pt,draw=white,fill=white]{$\mu$}
          --(0.7,0) node[circle,inner sep=0pt,draw=white,fill=white]{$V$}
          --(1.4,0) node[circle,inner sep=0pt,draw=white,fill=white]{$\nu$};
        \draw[thick]
            (0.7,0) node[circle,inner sep=0pt,draw=white,fill=white]{$V$}
          --(0.35,0.6) node[circle,inner sep=0pt,draw=white,fill=white]{$\sigma$};
        \draw[thick]
            (0.7,0) node[circle,inner sep=0pt,draw=white,fill=white]{$V$}
          --(1.05,0.6) node[circle,inner sep=0pt,draw=white,fill=white]{$\rho$};
      \end{tikzpicture}\vspace{-0.13cm}}
    =c_\alpha g^{\mu\rho}g^{\nu\sigma}+c_\beta g^{\mu\sigma}g^{\nu\rho}+c_\gamma g^{\mu\nu}g^{\sigma\rho},
\end{aligned}\end{equation}
where $c_\alpha$, $c_\beta$ and $c_\gamma$ are arbitrary coefficients. If we use one $\mathcal{O}(1)$ 4-vertex instead of two $\mathcal{O}(z)$ 3-vertices, one $\mathcal{O}(z^{-1})$ propagator will decrease. Basically, the more 4-vertices diagrammatic expressions have, the lower order they are. The Lorentz tensor $N_{\sigma(\mu_3\mu_4\cdots\mu_i)}$ can be expanded as
\begin{equation} \begin{aligned}
  N_{\sigma(\mu_3\mu_4\cdots\mu_i)}=
  N^{(0)}_{\sigma(\mu_3\mu_4\cdots\mu_i)}+\sum_{k=3}^{i-1}D_{\sigma(k)} N^{(1),\sigma(k)}_{\sigma(\mu_3\cdots\mu_i)}+\cdots,
\end{aligned} \end{equation}
where the superscript $(n)$ is the number of 4-vertices. Only the first two terms $N_0$ and $N_1$ (see figure \ref{npt}) give the same contributions in $\hat{A}_n^C$. Their diagrammatic expressions are
\begin{equation} \begin{aligned}
  N^{(0)}_{\mu_3\mu_4\cdots\mu_i}=\parbox[b]{4cm}{\HardNum{\hat{\epsilon}_1}{\hat{V}}{\hat{V}}{\hat{V}}{\hat{\epsilon}_2}{\mu_3}{\mu_4}{\mu_i}\vspace{-0.2cm}},
  \label{N0}
\end{aligned} \end{equation}
\begin{equation} \begin{aligned}
  N^{(1),k}_{\mu_3\cdots\mu_i}=
  \parbox[b]{4.7cm}{\HardNumfour{\hat{\epsilon}_1}{\hat{V}}{V}{\hat{V}}{\hat{\epsilon}_2}\vspace{-0.2cm}},
  \label{N1}
\end{aligned} \end{equation}
where the script $k$ denotes the position of the 4-vertex. Now we also evaluate these connected diagrammatic expressions in specific helicity and spin states.
\begin{equation} \begin{aligned}
  N^{(0),-,+}_{\mu_3\mu_4\cdots\mu_i}=&z^{i-3}\parbox[b]{4cm}{\HardNum{p_1}{R}{R}{R}{r}{\mu_3}{\mu_4}{\mu_i}\vspace{-0.2cm}}+\mathcal{O}(z^{i-4}),
\end{aligned} \end{equation}

\begin{equation} \begin{aligned}
  N^{(0),-,-}_{\mu_3\mu_4\cdots\mu_i}=&z^{i-2}C\parbox[b]{4cm}{\HardNum{p_1}{R}{R}{R}{p_1}{\mu_3}{\mu_4}{\mu_i}\vspace{-0.2cm}}
  -z^{i-3}\parbox[b]{0.3cm}{\Bigg[\vspace{-0.2cm}}
  \parbox[b]{4cm}{\HardNum{p_1}{R}{R}{R}{r^*}{\mu_3}{\mu_4}{\mu_i}\vspace{-0.2cm}}\\
  &-C\parbox[b]{0.3cm}{\Bigg(\vspace{-0.2cm}}
  \parbox[b]{4cm}{\HardNum{p_1}{V}{R}{R}{p_1}{\mu_3}{\mu_4}{\mu_i}\vspace{-0.2cm}}
  +\parbox[b]{4cm}{\HardNum{p_1}{R}{V}{R}{p_1}{\mu_3}{\mu_4}{\mu_i}\vspace{-0.2cm}}+\cdots\\
  &+\parbox[b]{4cm}{\HardNum{p_1}{R}{R}{V}{p_1}{\mu_3}{\mu_4}{\mu_i}\vspace{-0.2cm}}\parbox[b]{0.6cm}{\Bigg)\Bigg]\vspace{-0.2cm}}
  +\mathcal{O}(z^{i-4}),
\end{aligned} \end{equation}

\begin{equation} \begin{aligned}
  N^{(0),-,0}_{\mu_3\mu_4\cdots\mu_i}=&z^{i-2}C\parbox[b]{4cm}{\HardNum{p_1}{R}{R}{R}{r}{\mu_3}{\mu_4}{\mu_i}\vspace{-0.2cm}}
  -z^{i-3}\parbox[b]{0.3cm}{\Bigg[\vspace{-0.2cm}}
  \parbox[b]{4cm}{\HardNum{p_1}{R}{R}{R}{\bar{p}_2}{\mu_3}{\mu_4}{\mu_i}\vspace{-0.2cm}}\\
  &-C\parbox[b]{0.3cm}{\Bigg(\vspace{-0.2cm}}
  \parbox[b]{4cm}{\HardNum{p_1}{V}{R}{R}{r}{\mu_3}{\mu_4}{\mu_i}\vspace{-0.2cm}}
  +\parbox[b]{4cm}{\HardNum{p_1}{R}{V}{R}{r}{\mu_3}{\mu_4}{\mu_i}\vspace{-0.2cm}}+\cdots\\
  &+\parbox[b]{4cm}{\HardNum{p_1}{R}{R}{V}{r}{\mu_3}{\mu_4}{\mu_i}\vspace{-0.2cm}}\parbox[b]{0.6cm}{\Bigg)\Bigg]\vspace{-0.2cm}}
  +\mathcal{O}(z^{i-4}).
\end{aligned} \end{equation}
The leading diagrammatic expressions of $N^{(0)}$ vanish, because $r^2=(p_1 r)=0$. These numerators have simple tensor structures,
\begin{equation} \begin{aligned}
  N^{(0),-,-}_{\mu_3\cdots\mu_i}&=z^{i-3}C\sum_{a=3}^{i-1}\sum_{b=a+1}^{i} \left(\sum_{j=3}^{i} \alpha_j^{a,b}(p_j r)\right)  p_{1\mu_a} p_{1\mu_b} \prod_{l\neq a,b} r_{\mu_l}+\mathcal{O}(z^{i-4}),\\
  N^{(0),-,0}_{\mu_3\cdots\mu_i}&=z^{i-3}\sum_{a=3}^{i-1} \left(\sum_{j=3}^{i} \alpha_j^{a}(p_j r)\right)  p_{1\mu_a} \prod_{l\neq a} r_{\mu_l}+\mathcal{O}(z^{i-4}),
  \label{N0helicity}
\end{aligned} \end{equation}
where $\alpha_j^{a,b}$ and $\alpha_j^{a}$ are coefficients. In the subleading diagrammatic expressions of $N^{(0),-,-}$ and $N^{(0),-,0}$, $r^*$ and $\bar{p}_2$ only appear in the right end of these diagrams. We find that they have $i-2$ Lorentz indices and $i$ vectors, so there must be an inner product of two vectors. Since $r^2=(p_1 r)=(p_1 r^*)=(\bar{p}_2 r)=0$, this inner product should be
\begin{equation}\begin{aligned}
  \parbox[b]{4cm}{\HardNum{p_1}{R}{R}{R}{r^*}{\mu_3}{\mu_4}{\mu_i}\vspace{-0.2cm}}\propto(r^*r),\qquad 
  \parbox[b]{4cm}{\HardNum{p_1}{R}{R}{R}{\bar{p}_2}{\mu_3}{\mu_4}{\mu_i}\vspace{-0.2cm}}\propto(p_1\bar{p}_2).
\end{aligned}\end{equation}
The complete result of these expressions is given in eq.\,\eqref{MostR}. Then we use $(r^*r)+C(p_1p_2)=0$ and $(p_1p_2)=(p_1\bar{p}_2)$ to eliminate $r^*$ and $\bar{p}_2$. Eq.\,\eqref{N0helicity} also shows that $p_1$ should have free indices.

The leading diagrams of $N^{(0)}$ always vanish, but this is not true for $N^{(1)}$.
\begin{equation} \begin{aligned}
N^{(1),k,-,+}_{\mu_3\cdots\mu_i}=z^{i-5}\parbox[b]{4.7cm}{\HardNumfour{p_1}{R}{V}{R}{r}\vspace{-0cm}}+\mathcal{O}(z^{i-6}),
\end{aligned} \end{equation}

\begin{equation} \begin{aligned}
N^{(1),k,-,-}_{\mu_3\cdots\mu_i}
&=z^{i-4}C\parbox[b]{4.75cm}{\HardNumfour{p_1}{R}{V}{R}{p_1}\vspace{-0cm}}+\mathcal{O}(z^{i-5})\\
&=z^{i-4}C\sum_{a=3}^{i-1}\sum_{b=a+1}^{i}\beta_j^{a,b} p_1^{\mu_a} p_1^{\mu_b} \prod_{k\neq a,b} r^{\mu_k}+\mathcal{O}(z^{i-5}),\\
\end{aligned} \end{equation}

\begin{equation} \begin{aligned}
N^{(1),k,-,0}_{\mu_3\cdots\mu_i}
&=z^{i-4}\parbox[b]{4.7cm}{\HardNumfour{p_1}{R}{V}{R}{r}\vspace{-0cm}}+\mathcal{O}(z^{i-5})\\
&=z^{i-4}\sum_{a=3}^{i-1}\beta_j^{a} p_1^{\mu_a} \prod_{k\neq a} r^{\mu_k}+\mathcal{O}(z^{i-5}),\\
\end{aligned} \end{equation}
where $\beta_j^{a,b}$ and $\beta_j^{a}$ are coefficients. Besides these coefficients, we find that the leading contributions from $N^{(0)}$ and $N^{(1)}$ have same tensor structures.

Since the large-$z$ behavior of the denominator in the second term of eq.\,\eqref{connect1} is
\begin{equation} \begin{aligned}
  \prod_{j=3}^{i-1} D_{\sigma(j)}
  &=(2z)^{i-3}\prod_{j=3}^{i-1} \left(r\cdot\sum_{k=3}^j p_k\right)+\mathcal{O}(z^{i-4}),
\end{aligned} \end{equation}
For polarization $(h_1,s_2)=(-,-)$ and $(-,0)$, $N^{(0)}$ and $N^{(1)}$ give boundary terms in the large-$z$ limit. Therefore, the large-$z$ behavior of the connected part is
\begin{equation} \begin{aligned}
  \hat{A}_n^C\rightarrow \mathcal{O}(1),\quad n>3.
\end{aligned} \end{equation}

The disconnected diagrammatic expressions in $\hat{A}_n^D$ have more complicated structures than in $\hat{A}_n^C$. Since the second term in vector propagators disconnects the diagrammatic expressions, $\hat{A}_n^D$ should have at least one hard propagator,
\begin{equation} \begin{aligned}
  \hat{A}_n^D=&\sum_{\sigma\in S_{i-2}}\sum_{i=4}^{n}\frac{U_{\sigma(\mu_3\cdots\mu_i)}}{\prod_{j=3}^{i-1} D_{\sigma(j)}}\prod_{j=3}^{i} B^{\mu_j}_i,
  \label{disconnect1}
\end{aligned} \end{equation}
where $U_{\mu_3\cdots\mu_i}$ consists of three classes of Lorentz tensors, $\mathcal{L}$, $\mathcal{M}$ and $\mathcal{R}$.
\begin{equation} \begin{aligned}
  U_{\mu_3\cdots\mu_i}
  =&\sum_{3\le l<i}\mathcal{L}_{\mu_3\cdots\mu_{l}}\mathcal{R}_{\mu_{(l+1)}\cdots\mu_i}\\
  &+\sum_{m=1}^{i-4} \sum_{3\le l_1<l_2<\cdots<l_{m+1}<i} \mathcal{L}_{\mu_3\cdots\mu_{l_1}}\left(\prod_{a=1}^{m} \mathcal{M}_{\mu_{(l_a+1)}\cdots\mu_{l_{(a+1)}}}\right) \mathcal{R}_{\mu_{(l_{(m+1)}+1)}\cdots\mu_i},
  \label{disconnumer}
\end{aligned} \end{equation}
where $\mathcal{L}$ includes particle 1 and $\mathcal{R}$ includes particle 2. In analogy to $N_{\mu_3}$, Lorentz vectors $\mathcal{L}_{\mu_3}$, $\mathcal{M}_{\mu_l}$ and $\mathcal{R}_{\mu_i}$ don't include 4-vertices,
\begin{equation} \begin{aligned}
  \mathcal{L}_{\mu_3}&=
  \parbox[b]{1.9cm}{\Hards{\hat{\epsilon}_1}{\hat{V}}{\hat{P}}{\mu_3}\vspace{-0.15cm}}=p_3^2\hat{\epsilon}_{1\mu_3}-(p_3 \hat{\epsilon}_1)p_{3\mu_3}\rightarrow\mathcal{O}(z^{-1}),\\
  \mathcal{M}_{\mu_l}&=
  \parbox[b]{1.9cm}{\Hards{\hat{P}}{\hat{V}}{\hat{P}}{\mu_l}\vspace{-0.15cm}}=z\left(p_l^2 r_{\mu_l}-(p_l r)p_{l\mu_l}\right)+\mathcal{O}(1)\rightarrow\mathcal{O}(z),\\
  \mathcal{R}_{\mu_i}^+&=
  \parbox[b]{2cm}{\Hards{\hat{P}}{\hat{V}}{\hat{\epsilon}_2^+}{\mu_i}\vspace{-0.15cm}}\rightarrow\mathcal{O}(1),\\
  \mathcal{R}_{\mu_i}^0&=
  \parbox[b]{1.95cm}{\Hards{\hat{P}}{\hat{V}}{\hat{\epsilon}_2^0}{\mu_i}\vspace{-0.15cm}}=-2z^2 r_{\mu_i} r\cdot\sum_{j=3}^i p_j+\mathcal{O}(z)\rightarrow\mathcal{O}(z),\\
  \mathcal{R}_{\mu_i}^-&=
  \parbox[b]{2cm}{\Hards{\hat{P}}{\hat{V}}{\hat{\epsilon}_2^-}{\mu_i}\vspace{-0.15cm}}=-2z^2 p_{1\mu_i} r\cdot\sum_{j=3}^i p_j+\mathcal{O}(z)\rightarrow\mathcal{O}(z),
  \label{LMR1}
\end{aligned} \end{equation}
where the superscript of $\mathcal{R}_{\mu_i}$ denotes the polarization of particle 2. When $\mathcal{L}$, $\mathcal{M}$ and $\mathcal{R}$ have more than one Lorentz index, these tensors can be expanded in the number of 4-vertices as
\begin{equation} \begin{aligned}
  \mathcal{L}_{\mu_a\cdots\mu_b}&=\mathcal{L}^{(0)}_{\mu_a\cdots\mu_b}+\sum_{k=a}^{b-1} D_k \mathcal{L}^{(1),k}_{\mu_a\cdots\mu_b}+\cdots,\\
  \mathcal{M}_{\mu_a\cdots\mu_b}&=\mathcal{M}^{(0)}_{\mu_a\cdots\mu_b}+\sum_{k=a}^{b-1} D_k \mathcal{M}^{(1),k}_{\mu_a\cdots\mu_b}+\cdots,\\
  \mathcal{R}_{\mu_a\cdots\mu_b}&=\mathcal{R}^{(0)}_{\mu_a\cdots\mu_b}+\sum_{k=a}^{b-1} D_k \mathcal{R}^{(1),k}_{\mu_a\cdots\mu_b}+\cdots.\\
\end{aligned} \end{equation}

The diagrammatic expressions of $\mathcal{L}^{(0)}$, $\mathcal{M}^{(0)}$ and $\mathcal{R}^{(0)}$ are
\begin{equation} \begin{aligned}
  \mathcal{L}^{(0)}_{\mu_3 \cdots\mu_l}&=
  \parbox[b]{3.4cm}{\begin{tikzpicture}
    \draw[thick]
      --(0,0)node[circle,inner sep=0pt,draw=white,fill=white]{$\hat{\epsilon}_1^-$}
      --(1*0.7,0)node[circle,inner sep=0pt,draw=white,fill=white]{$\hat{V}$}
      --(2*0.7,0)node[circle,inner sep=1pt,draw=white,fill=white]{$\cdots$}
      --(3*0.7,0)node[circle,inner sep=0pt,draw=white,fill=white]{$\hat{V}$}
      --(4*0.7,0)node[circle,inner sep=0pt,draw=white,fill=white]{$\hat{P}$};
    \draw[thick]
        (0.7,0)node[circle,inner sep=0pt,draw=white,fill=white]{$\hat{V}$}
      --(0.7,0.7)node[circle,inner sep=0pt,draw=white,fill=white]{$\mu_3$};
    \draw[thick]
        (3*0.7,0)node[circle,inner sep=0pt,draw=white,fill=white]{$\hat{V}$}
      --(3*0.7,0.7)node[circle,inner sep=0pt,draw=white,fill=white]{$\mu_l$};
  \end{tikzpicture}\vspace{-0.23cm}}=N_{\mu_3\cdots\mu_l}^{-,0}+\mathcal{O}(z^{l-4}),
\end{aligned} \end{equation}

\begin{equation} \begin{aligned}
  \mathcal{M}^{(0)}_{\mu_{(l_a+1)}\cdots \mu_{l_{(a+1)}}}&=
  \parbox[b]{3.3cm}{\begin{tikzpicture}
    \draw[thick]
      --(0,0)node[circle,inner sep=0pt,draw=white,fill=white]{$\hat{P}$}
      --(1*0.7,0)node[circle,inner sep=0pt,draw=white,fill=white]{$\hat{V}$}
      --(2*0.7,0)node[circle,inner sep=1pt,draw=white,fill=white]{$\cdots$}
      --(3*0.7,0)node[circle,inner sep=0pt,draw=white,fill=white]{$\hat{V}$}
      --(4*0.7,0)node[circle,inner sep=0pt,draw=white,fill=white]{$\hat{P}$};
    \draw[thick]
        (0.7,0)node[circle,inner sep=0pt,draw=white,fill=white]{$\hat{V}$}
      --(0.7,0.7)node[circle,inner sep=-10pt,draw=white,fill=white]{$\mu_{(l_a+1)}$};
    \draw[thick]
        (3*0.7,0)node[circle,inner sep=0pt,draw=white,fill=white]{$\hat{V}$}
      --(3*0.7,0.7)node[circle,inner sep=-9pt,draw=white,fill=white]{$\mu_{l_{(a+1)}}$};
  \end{tikzpicture}\vspace{-0.23cm}}\\
  &=-2z^{l_{a+1}-l_a+1}\sum_{k=l_a+1}^{l_{a+1}-1}\sum_{j=3}^k(p_jr)\prod_{l} r_{\mu_l}+\mathcal{O}(z^{l_{a+1}-l_a}),
  \label{middle1}
\end{aligned} \end{equation}

\begin{equation} \begin{aligned}
  \mathcal{R}^{(0),s_2}_{\mu_{l+1}\cdots\mu_i}&=
  \parbox[b]{3.3cm}{\begin{tikzpicture}
    \draw[thick]
      --(0,0)node[circle,inner sep=0pt,draw=white,fill=white]{$\hat{P}$}
      --(1*0.7,0)node[circle,inner sep=0pt,draw=white,fill=white]{$\hat{V}$}
      --(2*0.7,0)node[circle,inner sep=1pt,draw=white,fill=white]{$\cdots$}
      --(3*0.7,0)node[circle,inner sep=0pt,draw=white,fill=white]{$\hat{V}$}
      --(4*0.7,0)node[circle,inner sep=-2pt,draw=white,fill=white]{$\hat{\epsilon}_2^{s_2}$};
    \draw[thick]
        (0.7,0)node[circle,inner sep=0pt,draw=white,fill=white]{$\hat{V}$}
      --(0.7,0.7)node[circle,inner sep=-5pt,draw=white,fill=white]{$\mu_{l+1}$};
    \draw[thick]
        (3*0.7,0)node[circle,inner sep=0pt,draw=white,fill=white]{$\hat{V}$}
      --(3*0.7,0.7)node[circle,inner sep=0pt,draw=white,fill=white]{$\mu_i$};
  \end{tikzpicture}\vspace{-0.23cm}}\rightarrow\mathcal{O}(z^{i-l+1}).
  \label{Rs2}
\end{aligned} \end{equation}
We find that $\mathcal{L}^{(0)}$ gives the same contribution as $N^{(0)}$ in the leading order. Since the order of $\mathcal{R}^{(0),+}$ in $z$ is lower than the other polarizations, we ignore $\mathcal{R}^{(0),+}$ in the following section. Here eq.\,\eqref{Rs2} is valid when $s_2=0$ or $+1$.

The diagrammatic expressions of $\mathcal{L}^{(1)}$, $\mathcal{M}^{(1)}$ and $\mathcal{R}^{(1)}$ are
\begin{equation} \begin{aligned}
  \mathcal{L}^{(1),k}_{\mu_3 \cdots\mu_l}=
  \parbox[b]{4.7cm}{\begin{tikzpicture}
  \draw[thick]
      (0,0)node[circle,inner sep=0pt,draw=white,fill=white]{$\hat{\epsilon}_1$}
    --(0.7,0)node[circle,inner sep=0pt,draw=white,fill=white]{$\hat{V}$}
    --(2*0.7,0)node[circle,inner sep=0pt,draw=white,fill=white]{$\cdots$}
    --(3*0.7,0)node[circle,inner sep=0pt,draw=white,fill=white]{$V$}
    --(4*0.7,0)node[circle,inner sep=0pt,draw=white,fill=white]{$\cdots$}
    --(5*0.7,0)node[circle,inner sep=0pt,draw=white,fill=white]{$\hat{V}$}
    --(6*0.7,0)node[circle,inner sep=0pt,draw=white,fill=white]{$\hat{P}$};
  \draw[thick]
      (0.7,0)node[circle,inner sep=0pt,draw=white,fill=white]{$\hat{V}$}
    --(0.7,0.7)node[circle,inner sep=0pt,draw=white,fill=white]{$\mu_3$};
  \draw[thick]
      (3*0.7-0.4,0.7)node[circle,inner sep=0pt,draw=white,fill=white]{$\mu_k$}
    --(3*0.7,0)node[circle,inner sep=0pt,draw=white,fill=white]{$V$}
    --(3*0.7+0.4,0.7)node[circle,inner sep=-4.5pt,draw=white,fill=white]{$\mu_{k+1}$};
  \draw[thick]
      (5*0.7,0)node[circle,inner sep=0pt,draw=white,fill=white]{$\hat{V}$}
    --(5*0.7,0.7)node[circle,inner sep=0pt,draw=white,fill=white]{$\mu_l$};
\end{tikzpicture}\vspace{-0.2cm}}=N^{(1),k}_{\mu_3 \cdots\mu_l}+\mathcal{O}(z^{l-5}),
\end{aligned} \end{equation}

\begin{equation} \begin{aligned}
  \mathcal{M}^{(1),k}_{\mu_{(l_a+1)}\cdots\mu_{l_{(a+1)}}}=
  \parbox[b]{4.7cm}{\begin{tikzpicture}
  \draw[thick]
      (0,0)node[circle,inner sep=0pt,draw=white,fill=white]{$\hat{P}$}
    --(0.7,0)node[circle,inner sep=0pt,draw=white,fill=white]{$\hat{V}$}
    --(2*0.7,0)node[circle,inner sep=0pt,draw=white,fill=white]{$\cdots$}
    --(3*0.7,0)node[circle,inner sep=0pt,draw=white,fill=white]{$V$}
    --(4*0.7,0)node[circle,inner sep=0pt,draw=white,fill=white]{$\cdots$}
    --(5*0.7,0)node[circle,inner sep=0pt,draw=white,fill=white]{$\hat{V}$}
    --(6*0.7,0)node[circle,inner sep=0pt,draw=white,fill=white]{$\hat{P}$};
  \draw[thick]
      (0.7,0)node[circle,inner sep=0pt,draw=white,fill=white]{$\hat{V}$}
    --(0.7,0.7)node[circle,inner sep=-10pt,draw=white,fill=white]{$\mu_{(l_a+1)}$};
  \draw[thick]
      (3*0.7-0.4,0.7)node[circle,inner sep=0pt,draw=white,fill=white]{$\mu_k$}
    --(3*0.7,0)node[circle,inner sep=0pt,draw=white,fill=white]{$V$}
    --(3*0.7+0.4,0.7)node[circle,inner sep=-4.5pt,draw=white,fill=white]{$\mu_{k+1}$};
  \draw[thick]
      (5*0.7,0)node[circle,inner sep=0pt,draw=white,fill=white]{$\hat{V}$}
    --(5*0.7,0.7)node[circle,inner sep=-9pt,draw=white,fill=white]{$\mu_{l_{(a+1)}}$};
\end{tikzpicture}\vspace{-0.2cm}}\rightarrow\mathcal{O}(z^{l_{a+1}-l_a}),
\end{aligned} \end{equation}

\begin{equation} \begin{aligned}
  \mathcal{R}^{(1),k}_{\mu_{l+1}\cdots\mu_i}=
  \parbox[b]{4.7cm}{\begin{tikzpicture}
  \draw[thick]
      (0,0)node[circle,inner sep=0pt,draw=white,fill=white]{$\hat{P}$}
    --(0.7,0)node[circle,inner sep=0pt,draw=white,fill=white]{$\hat{V}$}
    --(2*0.7,0)node[circle,inner sep=0pt,draw=white,fill=white]{$\cdots$}
    --(3*0.7,0)node[circle,inner sep=0pt,draw=white,fill=white]{$V$}
    --(4*0.7,0)node[circle,inner sep=0pt,draw=white,fill=white]{$\cdots$}
    --(5*0.7,0)node[circle,inner sep=0pt,draw=white,fill=white]{$\hat{V}$}
    --(6*0.7,0)node[circle,inner sep=0pt,draw=white,fill=white]{$\hat{\epsilon}_2$};
  \draw[thick]
      (0.7,0)node[circle,inner sep=0pt,draw=white,fill=white]{$\hat{V}$}
    --(0.7,0.7)node[circle,inner sep=-5pt,draw=white,fill=white]{$\mu_{l+1}$};
  \draw[thick]
      (3*0.7-0.4,0.7)node[circle,inner sep=0pt,draw=white,fill=white]{$\mu_k$}
    --(3*0.7,0)node[circle,inner sep=0pt,draw=white,fill=white]{$V$}
    --(3*0.7+0.4,0.7)node[circle,inner sep=-4.5pt,draw=white,fill=white]{$\mu_{k+1}$};
  \draw[thick]
      (5*0.7,0)node[circle,inner sep=0pt,draw=white,fill=white]{$\hat{V}$}
    --(5*0.7,0.7)node[circle,inner sep=0pt,draw=white,fill=white]{$\mu_i$};
\end{tikzpicture}\vspace{-0.2cm}}\rightarrow\mathcal{O}(z^{i-l}).
\end{aligned} \end{equation}

Combining the previous calculations, we find that the terms in $U_{\mu_3\cdots\mu_i}$ with more $\mathcal{M}$ generally give higher-order contributions. For example, consider
\begin{equation} \begin{aligned}
  \mathcal{M}_{\mu_a\cdots\mu_b}=\mathcal{O}(z^{b-a+2}).
  \label{middle2}
\end{aligned} \end{equation}
We use two $\mathcal{M}$ instead of it and find
\begin{equation} \begin{aligned}
  \mathcal{M}_{\mu_a\cdots\mu_c}\mathcal{M}_{\mu_{c+1}\cdots\mu_b}=\mathcal{O}(z^{c-a+2})\mathcal{O}(z^{b-c+1})=\mathcal{O}(z^{b-a+3}).
\end{aligned} \end{equation}
However, this rule doesn't work when $\mathcal{L}$, $\mathcal{M}$ and $\mathcal{R}$ have only one Lorentz index. Consider $\mathcal{M}_{\mu_a \mu_{a+1}}$, eq.\,\eqref{middle2} reduces to $\mathcal{O}(z^3)$, which gives the higher-order contribution than
\begin{equation} \begin{aligned}
  \mathcal{M}_{\mu_a}\mathcal{M}_{\mu_{a+1}}=\mathcal{O}(z)\mathcal{O}(z)=\mathcal{O}(z^2).
\end{aligned} \end{equation}
Therefore, the terms that have the most tensors with two Lorentz indices give the leading contributions. When $i>4$,
\begin{equation} \begin{aligned}
  U_{\mu_3\cdots\mu_i}
  &=\begin{cases}
    \mathcal{L}_{\mu_3}\left(\prod_{a=4}^{i-3}\mathcal{M}_{\mu_{a}\mu_{a+1}}\right)\mathcal{R}_{\mu_{i-1}\mu_i}+\cdots, &i=\mathrm{odd},\\
    \mathcal{L}_{\mu_3\mu_4}\left(\prod_{a=5}^{i-3}\mathcal{M}_{\mu_{a}\mu_{a+1}}\right)\mathcal{R}_{\mu_{i-1}\mu_i}+\cdots, &i=\mathrm{even},\\
  \end{cases}\\
  &=\begin{cases}
    \mathcal{O}(z^{-1})\left(\mathcal{O}(z^3)\right)^{\frac{i-5}{2}}\mathcal{O}(z^3), &i=\mathrm{odd},\\
    \mathcal{O}(z)\left(\mathcal{O}(z^3)\right)^{\frac{i-6}{2}}\mathcal{O}(z^3), &i=\mathrm{even},\\
  \end{cases}\\
  &=\begin{cases}
    \mathcal{O}(z^{\frac{3}{2}i-\frac{11}{2}}), &i=\mathrm{odd},\\
    \mathcal{O}(z^{\frac{3}{2}i-5}), &i=\mathrm{even}.\\
  \end{cases}
\end{aligned} \end{equation}
When $i=4$,
\begin{equation} \begin{aligned}
  U_{\mu_3\mu_4}=\mathcal{L}_{\mu_3}\mathcal{R}_{\mu_4}
  =\mathcal{O}(z^{-1})\mathcal{O}(z)=\mathcal{O}(1).
\end{aligned} \end{equation}
Substituting into eq.\,\eqref{disconnect1}, we get
\begin{equation} \begin{aligned}
  \hat{A}_n^D\rightarrow
  \begin{cases}
    \mathcal{O}(z^{-1}),  &n=4,\\
    \mathcal{O}(z^{\lfloor i/2\rfloor-2}), &n>4,\\
  \end{cases}
  \label{disconz}
\end{aligned} \end{equation}
where $\lfloor\cdots\rfloor$ is the floor function. It shows that the disconnected part doesn't vanish when $n>4$.

\subsection{Hard Scalars and Fermions}
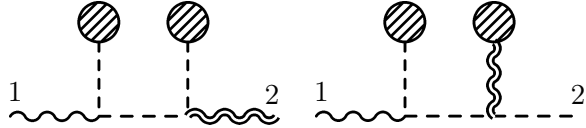
\begin{figure}[htbp]
  \centering
  \subfloat{
    \begin{fmfgraph*}(100,40)
      \fmfbottom{b1,b4}\fmfv{l=1,l.a=70,l.d=.06w}{b1}
      \fmf{photon}{b1,b2}
      \fmf{dashes}{b2,b3}
      \fmf{dbl_wiggly}{b3,b4}\fmfv{l=2,l.a=110,l.d=.05w}{b4}
      \fmffreeze
      \fmftopn{s}{4}
      \fmf{dashes}{s2,b2}
      \fmf{dashes}{s3,b3}
      \fmfblob{.15w}{s2,s3}
    \end{fmfgraph*}}\quad
  \subfloat{
    \begin{fmfgraph*}(100,40)
      \fmfbottom{b1,b4}\fmfv{l=1,l.a=70,l.d=.06w}{b1}
      \fmf{photon}{b1,b2}
      \fmf{dashes}{b2,b3}
      \fmf{dashes}{b3,b4}\fmfv{l=2,l.a=110,l.d=.05w}{b4}
      \fmffreeze
      \fmftopn{s}{4}
      \fmf{dashes}{s2,b2}
      \fmf{dbl_wiggly}{s3,b3}
      \fmfblob{.15w}{s2,s3}
    \end{fmfgraph*}}
  \caption{An example of two diagrams with hard scalar propagators.}
  \label{hardspropa}
\end{figure}

As we showed in section \ref{vertex}, the contributions from $VSS$ vertices are included in vector scattering amplitudes. Consider a diagrammatic expression in which two $VVV$ vertices are connected. Using eq.\,\eqref{substructure}, this expression can be expanded to
\begin{equation}
    \parbox[b]{2.6cm}{\Hardtu{\mu_1}{V}{V}{\mu_2}{\mu_3}{\mu_4}\vspace{-0.15cm}}=
    \parbox[b]{2.6cm}{\begin{tikzpicture}
      \draw[thick]
          (0.7,0.8)node[circle,inner sep=0pt,draw=white,fill=white]{$\mu_3$}
          --(0.7,0.25)--(1.4,0.25)
          --(1.4,0.8)node[circle,inner sep=0pt,draw=white,fill=white]{$\mu_4$};
      \draw[thick]
          (0,0)node[circle,inner sep=0pt,draw=white,fill=white]{$\mu_1$}
          --(0.7,0)node[circle,inner sep=0pt,draw=white,fill=white]{$V$};
      \draw[thick]
          (1.4,0)node[circle,inner sep=0pt,draw=white,fill=white]{$V$}
          --(2.1,0)node[circle,inner sep=0pt,draw=white,fill=white]{$\mu_2$};
      \end{tikzpicture}\vspace{-0.2cm}}+
    \parbox[b]{2.6cm}{\begin{tikzpicture}
      \draw[thick]
          (0.85,0.8)node[circle,inner sep=0pt,draw=white,fill=white]{$\mu_3$}
          --(0.85,0)--(1.4,0)
          --(2.1,0)node[circle,inner sep=0pt,draw=white,fill=white]{$\mu_2$};
      \draw[thick]
          (0,0)node[circle,inner sep=0pt,draw=white,fill=white]{$\mu_1$}
          --(0.6,0)node[circle,inner sep=0pt,draw=white,fill=white]{$V$};
      \draw[thick]
          (1.4,0.25)node[circle,inner sep=-.5pt,draw=white,fill=white]{$V$}
          --(1.4,0.8)node[circle,inner sep=-1pt,draw=white,fill=white]{$\mu_4$};
      \end{tikzpicture}\vspace{-0.2cm}}+\cdots,
\end{equation}
where $V^{\mu_1}V^{\mu_2}$ can be understood as two $VSS$ vertices, which correspond to the diagrams shown in figure \ref{hardspropa}. However, there is no $g^{\mu_3\mu_4}$ in the explicit expressions \eqref{N0helicity} and $p_1$ should contract with index $\mu_3$ or $\mu_4$. It means that the contributions from $VSS$ vertices vanish in the large-$z$ limit.

As discussed in refs. \cite{ArkaniHamed:2008yf,Cheung:2008dn}, the contributions from hard fermion propagators reduce to $\gamma^{\mu_j} \slashed{r} \gamma^{\mu_k}$, which can be viewed as a $\mathcal{O}(1)$ vertex. If we insert it into the vector amplitudes, the amplitudes will give lower-order contributions. We only need to evaluate the leading diagrammatic expressions of these contributions. Since
\begin{equation} \begin{aligned}
  r_{\mu_a} \gamma^{\mu_a} \slashed{r} \gamma^{\mu_b}=\gamma^{\mu_a} \slashed{r} \gamma^{\mu_b}r_{\mu_b}=0,\quad
  p_{1\mu_a} \gamma^{\mu_a} \slashed{r} \gamma^{\mu_b} p_{\mu_b}=0,\\
\end{aligned} \end{equation}
the leading diagrams vanish. Therefore, the hard fermion propagators improve the large-$z$ behavior.

If particle 2 is a massive fermion, the polarization of particle 2 should be changed. According to eq.\,\eqref{shiftedf}, the shift spinor of particle 2 factorizes in terms of a Weyl spinor $v^-_1$ and a little-group spinor $\xi^I=\langle 1\mathbf{2}^J\rangle$. Therefore, the leading diagram in the large-$z$ limit can also factorize into a product of $\xi^I$ and the leading diagram in the massless case. Using
\begin{equation} \begin{aligned}
  \slashed{r}v_1^-=p_2|1\rangle\langle11\rangle=0
\end{aligned} \end{equation}
and the Weyl equation $\slashed{p}_1v_1^-=0$, it is easy to check that the leading diagrams are equal to zero. Similarly, we can make the same conclusion on disconnected diagrams.

\section{Cancellation and Jacobi identity}
\label{sec5}
Since the amplitude with at least one gluon can be well recursed in the massless theory, we have reason to believe that this argument can be generalized into the massive theory. In last section, we evaluated all amplitudes with at least one massless vector boson in the large-$z$ limit. We found that they all vanish except in the vector scattering amplitudes.

Fortunately, the non-vanishing terms from $N^{(0)}_{\mu_3\cdots\mu_i}$ and $N^{(1)}_{k,\mu_3\cdots\mu_i}$ have similar structures in the large-$z$ limit, so there can be a cancellation among them.

\subsection{Cancellation in 4-point amplitudes}
\label{cancell4}
In the case of 4-point amplitudes, eq.\,\eqref{disconz} showed that the disconnected part $\hat{A}_4^D$ vanishes, so we just consider the connected part $\hat{A}_4^C$. Since the s-channel amplitude $N^{(0)}_{\mu_3}B^{\mu_3}_3$ vanishes, the total amplitude becomes
\begin{equation}
    \hat{A}_4^C=\left(\frac{N^{(0)}_{\mu_3\mu_4}}{D_3}+\frac{N^{(0)}_{\sigma_3(\mu_3\mu_4)}}{D_{\sigma_3(3)}}+N^{(1),3}_{\mu_3\mu_4}\right) B^{\mu_3}_4 B^{\mu_4}_4
    \rightarrow \mathcal{O}(z^{-1}),
    \label{amptotal}
\end{equation}
where $B_4^{\mu_3}=\epsilon_3^{\mu_3}$, $B_4^{\mu_4}=\epsilon_4^{\mu_4}$. The permutation $\sigma_3=(3\ 4)$ only exchanges particles $3$ and $4$.

Up until now, we haven't considered the group structure. Actually, the three-vector amplitudes \eqref{WWA} and \eqref{WWW} are not complete. We should introduce an antisymmetric tensor $f^{abc}$ to maintain Bose symmetry, so
\begin{equation}\begin{aligned}
    \parbox[b]{1.9cm}{\Hards{\mu_1}{V}{\mu_3}{\mu_2}\vspace{-0.15cm}}
    \rightarrow f^{a_1a_2a_3} V_{k,p,q}^{\mu_1\mu_2\mu_3}.
\end{aligned}\end{equation}
When $s_2=-1$, the diagrammatic expressions $N^{(0)}$ and $N^{(1)}$ reduce to
\begin{equation}\begin{aligned}
    N^{(0),-,-}_{\mu_3\mu_4} &=-2c_t Cz p_{1\mu_3} p_{1\mu_4}(p_3 r)+\mathcal{O}(1),\\
    N^{(0),-,-}_{\sigma_3(\mu_3\mu_4)} &=2c_u Cz p_{1\mu_3} p_{1\mu_4}(p_4 r)+\mathcal{O}(1),\\
    N^{(1),3,-,-}_{\mu_3\mu_4} &=(c_\alpha+c_\beta)Cp_{1\mu_3} p_{1\mu_4}+\mathcal{O}(z^{-1}),
    \label{ampmm}
\end{aligned}\end{equation}
where color factors $c_t=f^{a_1a_3b}f^{ba_4a_2}$, $-c_u=f^{a_1a_4b}f^{ba_3a_2}$. Substituting eq.\,\eqref{ampmm} into eq.\,\eqref{amptotal}, we derive a condition between color factors and 4-vertex coefficients, i.e. $c_\alpha+c_\beta=c_t-c_u$. Notice that, the particles 3 and 4 are vector bosons too. We can use the $[1,3\rangle$ and $[1,4\rangle$-shifts or their $\mathbf{BOLD}$ version to give extra conditions to validate eq.\,\eqref{amptotal}. The full conditions which ensure that the cancellation happens are
\begin{equation}
  \left\{
  \begin{aligned}
    c_\alpha+c_\beta &=c_t-c_u\\
    c_\alpha+c_\gamma &=c_u-c_s\\
    c_\beta+c_\gamma &=c_s-c_t,\\
  \end{aligned}\right.
\end{equation}
where $c_s=f^{a_1a_2e}f^{ea_3a_4}$.

This linear system has one unique solution, so the 4-vertex are determined by the constructibility of $A^{-,-}_{\mathrm{total}}$.
\begin{equation}
  \left\{
  \begin{aligned}
    c_\alpha &=c_t-c_s\\
    c_\beta &=c_s-c_u\\
    c_\gamma &=c_u-c_t,\\
  \end{aligned}\right.
\end{equation}

When $s_2=0$, the diagrammatic expressions $N^{(0)}$ and $N^{(1)}$ reduce to
\begin{equation}\begin{aligned}
    N^{(0),-,0}_{\mu_3\mu_4} &=-2c_t z(2r_{\mu_3}p_{1\mu_4}-p_{1\mu_3}r_{\mu_4})(p_3 r)+\mathcal{O}(z^{-1}),\\
    N^{(0),-,0}_{\sigma_3(\mu_3\mu_4)} &=-2c_u z(2r_{\mu_4}p_{1\mu_3}-p_{1\mu_4}r_{\mu_3})(p_4 r)+\mathcal{O}(z^{-1}),\\
    N^{(1),3,-,0}_{\mu_3\mu_4} &=c_\alpha r_{\mu_3}p_{1\mu_4}+c_\beta p_{1\mu_3} r_{\mu_4}+\mathcal{O}(z^{-1}).
  \label{ampmp}
\end{aligned}\end{equation}
Substituting eq.\,\eqref{ampmp} into eq.\,\eqref{amptotal} again, we derive another condition between color factor and 4-vertex coefficient,
\begin{equation}
  \left\{
  \begin{aligned}
    c_\alpha&=2c_t+c_u\\
    c_\beta&=-c_t-2c_u\\
  \end{aligned}\right.\rightarrow
  c_s+c_t+c_u=0.
\end{equation}
Finally, we derive the Jacobi identity in group theory. This implies that the constructibility of $A^{-,-}_{\mathrm{total}}$ and $A^{-,0}_{\mathrm{total}}$ can reconstruct a group structure in the massive amplitude. On the other hand, we can actually use recursion relations to construct the massive amplitude with gauge bosons. For example, the $n$-point amplitude involving 2 massive vector bosons and $n-2$ gluons has been explicitly constructed in ref. \cite{Ballav:2021ahg} under a massless-massive shift.

\subsection{Cancellation in \texorpdfstring{$n$}{n}-point amplitudes}
In section \ref{cancell4}, we determined the coefficients of 4-vertex,
\begin{equation}\begin{aligned}
    \parbox[b]{1.9cm}{
      \begin{tikzpicture}
        \draw[thick]
            (0,0) node[circle,inner sep=0pt,draw=white,fill=white]{$\mu_1$}
          --(0.7,0) node[circle,inner sep=0pt,draw=white,fill=white]{$V$}
          --(1.4,0) node[circle,inner sep=0pt,draw=white,fill=white]{$\mu_2$};
        \draw[thick]
            (0.7,0) node[circle,inner sep=0pt,draw=white,fill=white]{$V$}
          --(0.35,0.6) node[circle,inner sep=0pt,draw=white,fill=white]{$\mu_3$};
        \draw[thick]
            (0.7,0) node[circle,inner sep=0pt,draw=white,fill=white]{$V$}
          --(1.05,0.6) node[circle,inner sep=0pt,draw=white,fill=white]{$\mu_4$};
      \end{tikzpicture}\vspace{-0.13cm}}
    \rightarrow& (f^{a_1a_3e}f^{ea_4a_2}-f^{a_1a_2e}f^{ea_3a_4}) g^{\mu_1\mu_4}g^{\mu_2\mu_3}\\
    &+(f^{a_1a_2e}f^{ea_3a_4}+f^{a_1a_4e}f^{ea_3a_2})g^{\mu_1\mu_3}g^{\mu_2\mu_4}\\
    &-(f^{a_1a_4e}f^{ea_3a_2}+f^{a_1a_3e}f^{ea_4a_2})g^{\mu_1\mu_2}g^{\mu_3\mu_4},
\end{aligned}\end{equation}
and found that there is a cancellation in 4-point amplitudes, which is related to the group structure of massive vector bosons. Will this cancellation work on $n$-point amplitudes?

In the large-$z$ limit, the leading term in the denominator is a product of
\begin{equation} \begin{aligned}
  d_j=\sum_{k=3}^j(p_k r),\quad j=3,4,\cdots,i-1.
  \label{dpr}
\end{aligned} \end{equation}
In eq.\,\eqref{N0helicity}, we have shown that there are some inner products $(p_k r)$, which don't occur in $N^{(1)}$. Therefore, we invert eq.\,\eqref{dpr} to give $(p_j r)$ in terms of $d_k$,
\begin{equation} \begin{aligned}
  (p_j r)=\left\{\begin{aligned}
    &d_j, &j&=3,\\
    &d_j-d_{j-1}, &3<j&<i,\\
    &-d_{j-1}, &j&=i.\\
  \end{aligned}\right.
\end{aligned} \end{equation}
Now we choose a basis $\{d_k\}$ and consider the decomposition of $N^{(0)}$,
\begin{equation} \begin{aligned}
  N^{(0)}_{\mu_3\cdots\mu_i}=\sum_{k=3}^{i-1} d_k N^{(0),k}_{\mu_3\cdots\mu_i}.
\end{aligned} \end{equation}
After the decomposition, we collect the contributions from $N^{(0)}$ and $N^{(1)}$ that have same tensor structures, and define a new tensor
\begin{equation} \begin{aligned}
  N^k_{\mu_3\cdots\mu_i}=\frac{1}{2}\left(
  N^{(0),k}_{\mu_3\cdots\mu_i}+
  N^{(0),\sigma_k(k)}_{\sigma_k(\mu_3\cdots\mu_i)}+
  N^{(1),k}_{\mu_3\cdots\mu_i}\right).
\end{aligned} \end{equation}
where the permutation $\sigma_k=(k\ k+1)$ is an element of symmetric group $S_{i-2}$ and the factor $1/2$ is introduced to avoid double counting. Appendix \ref{AppC} gives an example that explicitly evaluates this tensor and corresponding diagrammatic expressions.

In $N^k$, the $p_1$ from $\epsilon_1$ only contracts with the index $\mu_k$ or $\mu_{k+1}$. This structure is contained in an antisymmetric tensor
\begin{equation} \begin{aligned}
  \mathcal{P}^{k,k+1}=p_{1\mu_k}r_{\mu_{k+1}}-r_{\mu_k}p_{1\mu_{k+1}}.
\end{aligned} \end{equation}
Now the $N^k$ in specific polarizations becomes
\begin{equation} \begin{aligned}
  N^{k,-,-}_{\mu_3\cdots\mu_i}=z^{i-3}\frac{C}{2}(c_1-c_2+c_3)\mathcal{P}^{k,k+1}\sum_{j\neq k,k+1} F^{i,j,k} p_{1\mu_j}\prod_{l\neq j,k,k+1} r_{\mu_l}+\mathcal{O}(z^{i-4}),
  \label{canceln1}
\end{aligned} \end{equation}

\begin{equation} \begin{aligned}
  N^{k,-,0}_{\mu_3\cdots\mu_i}=z^{i-3}(c_1-c_2+c_3)\mathcal{P}^{k,k+1}2^{k-4}\prod_{j\neq k,k+1} r_{\mu_j}+\mathcal{O}(z^{i-4}),
  \label{canceln2}
\end{aligned} \end{equation}
where $F^{i,j,k}$ is
\begin{equation} \begin{aligned}
  F^{i,j,k}=\left\{ \begin{aligned}
    &2^{i-4+j-k}, &1\le j<k,\\
    &2^{i-3+k-j}, &k+1<j\le i.\\
  \end{aligned} \right.
\end{aligned} \end{equation}

The color factors in eqs. \eqref{canceln1} and \eqref{canceln2} are
\begin{equation} \begin{aligned}
  c_1&=f^{a_1 a_3 b_1}\cdots f^{b_{k-2} a_k b_{k-1}}f^{b_{k-1} a_{k+1} b_k} \cdots f^{b_1 a_n a_2},\\
  c_2&=f^{a_1 a_3 b_1}\cdots f^{b_{k-2} a_{k+1} b_{k-1}}f^{b_{k-1} a_{k} b_k} \cdots f^{b_1 a_n a_2},\\
  c_3&=f^{a_1 a_3 b_1}\cdots (f^{b_{k-2} e b_k}f^{a_k e a_{k+1}}) \cdots f^{b_1 a_n a_2}.
\end{aligned} \end{equation}
If these factors have a cancellation
\begin{equation} \begin{aligned}
  c_1-c_2+c_3=0,
\end{aligned} \end{equation}
the large-$z$ behavior of connected diagrammatic expressions will be improved,
\begin{equation} \begin{aligned}
  N^{k}_{\mu_3\cdots\mu_i}\rightarrow \mathcal{O}(z^{i-4}).
\end{aligned} \end{equation}

Using $N^{k}$ and eq.\,\eqref{capitald}, we can rewrite the leading contribution of $\hat{A}^C_n$,
\begin{equation} \begin{aligned}
  \hat{A}^C_n=\sum_{i=4}^n\sum_{\sigma\in S_{i-2}}
  \frac{\sum_{k=3}^{i-1}d_{\sigma(k)} N^{\sigma(k)}_{\sigma(\mu_3\cdots\mu_i)}}{(2z)^{i-3}\prod_{j=3}^{i-1}\sigma(d_j)}\prod_{j=3}^{i}B^{\mu_j}_i+\mathcal{O}(z^{-1}).
\end{aligned} \end{equation}
After the cancellation, we find that $\hat{A}^C_n$ vanishes in the large-$z$ limit.

We can also decompose the disconnected diagrammatic expressions
\begin{equation} \begin{aligned}
  \mathcal{L}_{\mu_3 \cdots\mu_l}&=
  \sum_{k=3}^{l-1} d_k \mathcal{L}_{\mu_3 \cdots\mu_l}^{k},\\
  \mathcal{M}_{\mu_{(l_1+1)}\cdots\mu_{l_{2}}}&=
  \sum_{k=l_1+1}^{l_{2}-1} d_k \mathcal{M}^{k}_{\mu_{(l_1+1)}\cdots\mu_{l_{2}}},\\
  \mathcal{R}_{\mu_{l+1} \cdots\mu_i}&=
  \sum_{k=l+1}^{i-1} d_k \mathcal{R}^{k}_{\mu_{l+1} \cdots\mu_i},
\end{aligned} \end{equation}
where
\begin{equation} \begin{aligned}
  \mathcal{L}_{\mu_3 \cdots\mu_l}^{k}&=\frac{1}{2}\left(\mathcal{L}^{(0),k}_{\mu_3 \cdots\mu_l}-\mathcal{L}^{(0),\sigma(k)}_{\sigma_k(\mu_3 \cdots\mu_l)}+\mathcal{L}^{(1),k}_{\mu_3 \cdots\mu_l}\right),\\
  \mathcal{M}_{\mu_{(l_1+1)}\cdots\mu_{l_{2}}}^{k}&=\frac{1}{2}\left(\mathcal{M}^{(0),k}_{\mu_{(l_1+1)}\cdots\mu_{l_{2}}}-\mathcal{M}^{(0),\sigma_k(k)}_{\sigma_k(\mu_{(l_1+1)}\cdots\mu_{l_{2}})}+\mathcal{M}^{(1),k}_{\mu_{(l_1+1)}\cdots\mu_{l_{2}}}\right),\\
  \mathcal{R}^{k}_{\mu_{l+1} \cdots\mu_i}&=\frac{1}{2}\left(\mathcal{R}^{(0),k}_{\mu_{l+1} \cdots\mu_i}-\mathcal{R}^{(0),\sigma_k(k)}_{\sigma_k(\mu_{l+1} \cdots\mu_i)}+\mathcal{R}^{(1),k}_{\mu_{l+1} \cdots\mu_i}\right).
\end{aligned} \end{equation}
They are the analog of $N^k$. Since $\mathcal{L}$ gives the same leading contributions as $N^{-,0}$, we don't need to evaluate it again. As we showed in eq.\,\eqref{middle1}, $\mathcal{M}^{(0)}$ has simple tensor structure and will cancel $\mathcal{M}^{(1)}$ directly. The last expression $\mathcal{R}$ gives
\begin{equation} \begin{aligned}
  \mathcal{R}^{k,0}_{\mu_{l+1}\cdots\mu_i}&=\mathcal{O}(z^{i-l}),\\
  \mathcal{R}^{k,-}_{\mu_{l+1}\cdots\mu_i}&=\frac{1}{2}(c_1-c_2+c_3)\mathcal{P}^{k,k+1}\prod_{j\neq k,k+1} r_{\mu_j}+\mathcal{O}(z^{i-l}),
\end{aligned} \end{equation}
where the $p_1$ in $\mathcal{P}^{k,k+1}$ comes from $\epsilon_2^-$. When $s_2=0$, we use $r$ instead of $p_1$, so $\mathcal{P}^{k,k+1}$ gives zero.

After the cancellation $(c_1-c_2+c_3)=0$, we summarize the large-$z$ behavior of the tensor $\mathcal{L}$, $\mathcal{M}$ and $\mathcal{R}$ with one and more than one Lorentz index,
\begin{equation} \begin{aligned}
  \mathcal{L}_{\mu_3 \cdots\mu_l}\rightarrow \mathcal{O}(z^{l-4}),\quad
  \mathcal{M}_{\mu_{(l_a+1)}\cdots\mu_{l_{(a+1)}}}\rightarrow
  \mathcal{O}(z^{l_{a+1}-l_a}),\quad
  \mathcal{R}_{\mu_{l+1} \cdots\mu_i}\rightarrow
  \mathcal{O}(z^{i-l}).
\end{aligned} \end{equation}
We use these building blocks to calculate the two terms in eq.\,\eqref{disconnumer},
\begin{equation} \begin{aligned}
  L_{\mu_3\cdots\mu_l} R_{\mu_{l+1}\cdots\mu_i}
  \rightarrow \mathcal{O}(z^{l-4})\mathcal{O}(z^{i-l})=\mathcal{O}(z^{i-4}),
\end{aligned} \end{equation}

\begin{equation} \begin{aligned}
  &L_{\mu_3\cdots\mu_{l_1}}\left(\prod_{a=1}^{m} M_{\mu_{(l_a+1)}\cdots\mu_{l_{(a+1)}}}\right) R_{\mu_{(l_{(m+1)}+1)}\cdots\mu_i}\\
  &\rightarrow\mathcal{O}(z^{l_1-4})\left(\prod_{a=1}^{m}\mathcal{O}(z^{l_{a+1}-l_a})\right)\mathcal{O}(z^{i-l_{m+1}})=\mathcal{O}(z^{i-4}).
\end{aligned} \end{equation}
It shows that the large-$z$ behavior of disconnected diagrammatic expressions are also improved. Therefore, the disconnected part vanishes in the large-$z$ limit.

\subsection{4-point all-massive amplitudes}

\begin{table}[ht]
  \begin{center}
    \begin{tabular}{|c|c|c|c|}
      \hline
      $s_1\backslash s_2$ & $-1$ & $0$ & $1$ \\ \hline
      $-1$ & $\mathcal{O}(z)$ & $\mathcal{O}(1)$ & $\mathcal{O}(z^{-1})$ \\ \hline
      $0$ & $\mathcal{O}(z^2)$ & $\mathcal{O}(z)$ & $\mathcal{O}(1)$ \\ \hline
      $1$ & $\mathcal{O}(z^3)$ & $\mathcal{O}(z^2)$ & $\mathcal{O}(z)$ \\ \hline
    \end{tabular}
  \end{center}
  \caption{The large-$z$ behavior of amplitudes with particles 1 and 2 in different spin states. $s_1$ and $s_2$ are spin projections of particles 1 and 2 in the $z$ direction.}
  \label{fourmassive}
\end{table}

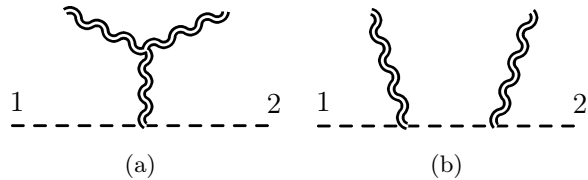
\begin{figure}[htbp]
  \centering
    \subfloat[]{\label{goldstone1}
      \begin{fmfgraph*}(100,50)
        \fmfbottom{i1,o1}\fmfv{l=1,l.a=70,l.d=.06w}{i1}
        \fmf{dashes}{i1,v1}
        \fmf{dashes}{v1,o1}\fmfv{l=2,l.a=110,l.d=.05w}{o1}
        \fmffreeze
        \fmftopn{s}{6}
        \fmf{dbl_wiggly}{v1,v2}
        \fmf{dbl_wiggly}{s2,v2}
        \fmf{dbl_wiggly}{v2,s5}
      \end{fmfgraph*}}\quad
    \subfloat[]{\label{goldstone2}
      \begin{fmfgraph*}(100,50)
        \fmfbottom{i1,o1}
        \fmf{dashes}{i1,v1}\fmfv{l=1,l.a=70,l.d=.06w}{i1}
        \fmf{dashes}{v1,v2}
        \fmf{dashes}{v2,o1}\fmfv{l=2,l.a=110,l.d=.05w}{o1}
        \fmffreeze
        \fmftopn{s}{6}
        \fmf{dbl_wiggly}{s2,v1}
        \fmf{dbl_wiggly}{v2,s5}
      \end{fmfgraph*}}
  \caption{Amplitudes with hard particles replaced by Goldstone bosons}
  \label{goldrepla}
\end{figure}

One may wonder whether the above cancellation happens in the amplitude with all vectors massive. Consider a 4-point amplitude under the $[\mathbf{1},\mathbf{2}\rangle$-shifts, both Higgs exchange and vector exchange diagrams will contribute,
\begin{equation} \begin{aligned}
    A_{\mathrm{total}}=A_{V}+A_{\mathrm{Higgs}}\propto[c_s A_{\ref*{goldstone1}}+(c_t+c_u)A_{\ref*{goldstone2}}].
\end{aligned} \end{equation}
Table \ref{fourmassive} shows the large-$z$ behavior of the amplitude in different spin states. We find that only $t$ and $u$-channel $A_{\mathrm{Higgs}}$ amplitudes will give the same order as shown in table \ref{fourmassive}. Suppose the coupling of massive vectors and Higgs is $m$, the leading contribution is
\begin{equation} \begin{aligned}
    A_{\mathrm{Higgs}}=-m^2\left(c_t\frac{(\hat{\epsilon}_1\epsilon_3)(\epsilon_4\hat{\epsilon}_2)}{2(p_3 r)}+c_u\frac{(\hat{\epsilon}_1\epsilon_4)(\epsilon_3\hat{\epsilon}_2)}{2(p_4 r)}\right)+\cdots.
\end{aligned} \end{equation}

Then we consider $A_{V}$. Since particle 1 is no longer a massless gauge boson, $\epsilon_1\rightarrow -p_1/z$ cannot be used in calculations. We take $(s_1,s_2)=(-,-)$ as an example and substitute polarizations \eqref{polar3} into the shifted amplitude, and the numerator becomes
\begin{equation}
  \begin{aligned}
    N^{(0),-,-}_{\mu_3\mu_4}=&-z^3
    \parbox[b]{2.6cm}{\Hardtu{r}{R}{R}{\tilde{p}_1}{\mu_3}{\mu_4}\vspace{-0.16cm}}\\
    &+z^2\parbox[b]{0.3cm}{\Bigg[\vspace{-0.2cm}}
    \parbox[b]{2.5cm}{\Hardtu{r}{R}{R}{r^*}{\mu_3}{\mu_4}\vspace{-0.14cm}}-\parbox[b]{0.3cm}{\Bigg(\vspace{-0.2cm}}
    \parbox[b]{2.5cm}{\Hardtu{r}{R}{V}{\tilde{p}_1}{\mu_3}{\mu_4}\vspace{-0.17cm}}+
    \parbox[b]{2.5cm}{\Hardtu{r}{V}{R}{\tilde{p}_1}{\mu_3}{\mu_4}\vspace{-0.17cm}}\parbox[b]{0.3cm}{\Bigg)\vspace{-0.2cm}}\parbox[b]{0.3cm}{\Bigg]\vspace{-0.2cm}}+\mathcal{O}(z).\\
  \end{aligned}
\end{equation}
The leading diagram is still equal to zero. However, the subleading diagram will contribute a new term proportional to $p_1^2$, which cannot be canceled with the contribution from the 4-vertex or the $s$-channel. Similarly, the subleading diagrams with particles 1 and 2 in other spin states may contribute non-vanishing terms.

Combining these two contributions, we get
\begin{equation} \begin{aligned}
    A^{-,+}_{\mathrm{total}}&=-\frac{1}{z}\frac{2(p_1p_2)+m^2}{2}\tilde{A}+\mathcal{O}\left(z^{-2}\right),\\
    A^{-,0}_{\mathrm{total}}&=\frac{2(p_1p_2)-2(p_1\bar{p}_2)+m^2}{2}\tilde{A}+\mathcal{O}\left(z^{-1}\right),\\
    A^{-,-}_{\mathrm{total}}&=zC(p_1\tilde{p}_1)\tilde{A}+\mathcal{O}(1),\\
    A^{0,0}_{\mathrm{total}}&=z\frac{2(p_1p_2)+2(\bar{p}_1\bar{p}_2)-2(p_1\bar{p}_2)-2(\bar{p}_1p_2)+m^2}{2}\tilde{A}+\mathcal{O}(1),\\
    A^{0,-}_{\mathrm{total}}&=z^2C[(p_1\tilde{p}_1)-(\bar{p}_1\tilde{p}_1)]\tilde{A}+\mathcal{O}(z),\\
    A^{+,-}_{\mathrm{total}}&=-z^3C^2(\tilde{p}_1\tilde{p}_2)\tilde{A}+\mathcal{O}(z^2),
    \label{fourma}
\end{aligned} \end{equation}
where $\tilde{p}_1=C(p_1+\bar{p}_1)/2$ and $\tilde{p}_2=C(p_2+\bar{p}_2)/2$. Interestingly, all the leading contributions in eq.\,\eqref{fourma} are proportional to $\tilde{A}$, which also contribute to the amplitudes with hard particles replaced by Goldstone bosons (see figure \ref{goldrepla}).

After the replacement, the amplitude becomes
\begin{equation} \begin{aligned}
    A_{\mathrm{Goldstone}}&=z\tilde{A}+\mathcal{O}(1),
    \label{goldstone}
\end{aligned} \end{equation}
where
\begin{equation} \begin{aligned}
    \tilde{A}=4c_s\frac{(p_4\epsilon_3)(r\epsilon_4)-(p_3\epsilon_4)(r\epsilon_3)+(p_3r)(\epsilon_3\epsilon_4)}{s-m^2}+2(c_t+c_u)\frac{(r\epsilon_3)(r\epsilon_4)}{(p_3r)}.
\end{aligned} \end{equation}

It implies that the 4-point amplitudes with all vectors massive have some structure other than gauge structure, which can be described by introducing scalar bosons. In the high energy limit, this structure concentrates on the longitudinal mode of massive vectors and becomes what is called Goldstone boson equivalence theorem. In two-line complex deformations, this structure appears in all polarization compositions.

There have been some works \cite{Feng:2009ei,Feng:2010ku} on how to compute the boundary term $B_n$ systematically. If we want to use these methods to evaluate the boundary term in the amplitude with all vectors massive, it must be more complicated than the massless case.

\section{Conclusions and discussions}
  \label{sec6}
  We have constructed the complete two and three-line shifts for all masses and provided a method to construct general multi-line shifts. These various shifts may be useful in investigating massive amplitude structures. For example, soft $\epsilon$-shift \cite{Cheung:2015cba} and BCFW shifts can be combined to give a two-parameter shift \cite{Elvang:2016qvq}, which is useful to explore soft theorems in the massless case. Although there have been some effort to construct soft recursions in the massive case \cite{Falkowski:2020aso}, our research provides a possibility to construct other useful soft shifts in the future.

  The validity of massless-massive BCFW shifts have been examined in the $n$-point amplitudes in this paper. For a theory of spin $\le 1$, any amplitude with at least one massless gauge boson can be recursive constructed. Furthermore, we can use constructibility to derive the group structure for the amplitude with massless and massive vectors. Since gluon amplitudes also have group structures, the massless gauge boson can be naturally viewed as the transverse component of a massive vector boson in the higher energy limit. For the amplitude with all vectors massive, we found that not only longitudinal mode but also transverse mode of massive vectors contribute similar amplitude structures as Goldstone bosons in the large-$z$ limit. Conversely, we can set the massive amplitudes proportional to the amplitudes replaced by Goldstone in the large-$z$ limit, so that we can determine the coupling of Higgs and massive vectors.

  Though the masses of vector bosons are equal in our analysis, the conclusion can be generalized to amplitudes in more general theories, e.g. the Standard Model and some BSM models. Since we can set up a specific gauge group and vacuum expectation value (VEV), vector bosons will have different masses after symmetry breaking. This manipulation only changes the coefficients in amplitudes in the large-$z$ limit, so our result is still valid. The success of generalization in this paper encourages us to consider more general cases in massive amplitudes. It is interesting to see whether our analysis can be generalized to theories of spin $\le 2$ when massive particles are included.

  \section*{Acknowledgements}
  CW thanks Teng Ma, Jing Shu, Ming-Lei Xiao and Yu-Hui Zheng for the useful discussions. This work was supported by the National Science Foundation of China under Grants No. 11875072, 11635001.

\appendix

\section{Massless and massive spinor-helicity formalism}
  \label{AppA}

  \setcounter{equation}{0}
  \renewcommand{\theequation}{A.\arabic{equation}}

  For a massless particle, whose momentum is
  \begin{equation}
    p^\mu =(E,E\sin\theta\cos\phi,E\sin\theta\sin\phi,E\cos\theta).
  \end{equation}
  The momentum multiplied by gamma matrices factorizes into a product of two-component Weyl spinors,
  \begin{equation} \begin{aligned}
    p_{\alpha\dot{\beta}}=p_\mu \sigma^\mu_{\alpha\dot{\beta}}=|p]_\alpha\langle p|_{\dot{\beta}}.\\
  \end{aligned} \end{equation}

  We choose the explicit form of these variables to be
  \begin{equation} \begin{aligned}
    |p\rangle^{\dot{\alpha}}=\sqrt{2E}\begin{pmatrix}
      c \\
      s \\
    \end{pmatrix},\langle p|_{\dot{\alpha}}=\sqrt{2E}\begin{pmatrix}
      -s & c \\
    \end{pmatrix},\\
    |p]_\alpha=\sqrt{2E}\begin{pmatrix}
      -s^* \\
      c \\
    \end{pmatrix},[p|^\alpha=\sqrt{2E}\begin{pmatrix}
      c & s^* \\
    \end{pmatrix},\\
    \label{spinormassless}
  \end{aligned} \end{equation}
  where $c=\cos(\theta/2)$, $s=\sin(\theta/2)\exp(i\phi)$. All Lorentz invariant structures can be expressed in terms of these variables:
  \begin{equation} \begin{aligned}
    \langle ij\rangle&=\langle i|_{\dot{\alpha}} |j\rangle^{\dot{\alpha}}=\epsilon^{\dot{\alpha}\dot{\beta}}\langle i|_{\dot{\alpha}} \langle j|_{\dot{\beta}},\\
    [ij]&=[i|_\alpha |j]^\alpha=\epsilon^{\alpha\beta}[i|_\alpha [j|_\beta,\\
  \end{aligned} \end{equation}
  where
  \begin{equation} \begin{aligned}
      \epsilon^{\alpha\beta}= \epsilon^{\dot{\alpha}\dot{\beta}}=-\epsilon_{\alpha\beta}=-\epsilon_{\dot{\alpha}\dot{\beta}}=
      \begin{pmatrix}
        0 & 1 \\
        -1 & 0 \\
      \end{pmatrix}.\\
    \end{aligned} \end{equation}
  Furthermore, Lorentz vectors can also be denoted by angle-square brackets:
  \begin{equation} \begin{aligned}
    \langle i|\gamma^\mu|j]=\langle i|_{\dot{\alpha}}\bar{\sigma}^{\mu\dot{\alpha}\beta}|j]_\beta=[ j|^{\alpha}\bar{\sigma}^\mu_{\alpha\dot{\beta}}|i\rangle^{\dot{\beta}}=[j|\gamma^\mu|i\rangle.
  \end{aligned} \end{equation}
  If $i=j$, it reduces to $\langle i|\gamma^\mu|i]=2p_i^\mu$.

  Now consider a massive particle whose momentum is
  \begin{equation}
    p^\mu =(E,p\sin\theta\cos\phi,p\sin\theta\sin\phi,p\cos\theta).
  \end{equation}
  The massive spinor variables of this particle are
  \begin{equation}
    \begin{aligned}
      |\mathbf{p}^I]_\alpha&=
      \begin{pmatrix}
        -\sqrt{E+p}s^* & \sqrt{E-p}c\\
        \sqrt{E+p}c & \sqrt{E-p}s
      \end{pmatrix},\\
      |\mathbf{p}^I\rangle^{\dot{\alpha}}&=
      \begin{pmatrix}
        \sqrt{E-p}s^* & -\sqrt{E+p}c\\
        -\sqrt{E-p}c & -\sqrt{E+p}s
      \end{pmatrix}.\\
      \end{aligned}
    \end{equation}
    where $I=1,2$ is little-group index. Here our conventions are different from \cite{Arkani-Hamed:2017jhn,Christensen:2018zcq}. In the high energy limit, these spinor variables reduce to eq.\,\eqref{spinormassless}. Contracting Weyl-spinor indices, we obtain
    \begin{equation}
      \begin{aligned}
        \langle\mathbf{p}_I\mathbf{p}_J\rangle &= m\epsilon_{IJ}, \\
        \left[\mathbf{p}_I\mathbf{p}_J\right] &= -m\epsilon_{IJ}.
      \end{aligned}
    \end{equation}
    where
    \begin{equation}
      \begin{aligned}
        \epsilon^{IJ}=-\epsilon_{IJ}=
        \begin{pmatrix}
          0 & 1 \\
          -1 & 0 \\
        \end{pmatrix}.\\
      \end{aligned}
    \end{equation}
    Contracting little-group indices, we obtain
    \begin{equation}
      \begin{aligned}
      p_{\alpha\dot{\beta}}=|\mathbf{p}^I]_\alpha\langle\mathbf{p}_I|_{\dot{\beta}}, \\
      p^{\dot{\alpha}\beta}=|\mathbf{p}_I\rangle^{\dot{\alpha}}[\mathbf{p}^I|^\beta.
      \end{aligned}
    \end{equation}

\section{Spinors in the two particles center-of-mass frame}
  \label{AppB}

  \setcounter{equation}{0}
  \renewcommand{\theequation}{B.\arabic{equation}}

  We take particles 1 and 2 to be massless and massive respectively, and choose the center-of-mass frame of these particles. The momenta of these particles become

  \begin{equation}
    \begin{aligned}
      p_1=(p,0,0,p),\quad p_2=(E,0,0,-p).
    \end{aligned}
  \end{equation}

  We define a null vector $\bar{p}_1^\mu=(p,0,0,-p)$, so $\langle\bar{1}1\rangle=[1\bar{1}]=-2p$. The massive spinors can be expanded by the spinor from $p_1$ and $\bar{p}_1$,

  \begin{equation}
    \begin{aligned}
      |\mathbf{2}^1]&=\sqrt{E+p}
      \begin{pmatrix}
        1\\
        0
      \end{pmatrix}=\sqrt{\frac{E+p}{2p}}|\bar{1}],
      &|\mathbf{2}^2]&=-\sqrt{E-p}
      \begin{pmatrix}
        0\\
        1
      \end{pmatrix}=-\sqrt{\frac{E-p}{2p}}|1],\\
      |\mathbf{2}^2\rangle&=\sqrt{E+p}
      \begin{pmatrix}
        0\\
        1
      \end{pmatrix}=-\sqrt{\frac{E+p}{2p}}|\bar{1}\rangle,
      &|\mathbf{2}^1\rangle&=-\sqrt{E-p}
      \begin{pmatrix}
        1\\
        0
      \end{pmatrix}=-\sqrt{\frac{E-p}{2p}}|1\rangle.\\
      \end{aligned}
    \end{equation}
  In the high energy limit, $|\mathbf{2}^1]$ and $|\mathbf{2}^2\rangle$ exist while $|\mathbf{2}^2]$ and $|\mathbf{2}^1\rangle$ vanish, so $[\mathbf{2}^2 1]=\langle\mathbf{2}^1 1\rangle=0$, $\langle\mathbf{2}^2 1\rangle=-\sqrt{2p(E+p)}$.

  Now we consider the $[1,\mathbf{2}\rangle$ shift \eqref{ij2} and rescale it,
  \begin{equation}
    |\hat{1}]=|1]+z|\mathbf{2}^I]\frac{\langle\mathbf{2}_I1\rangle}{E+p},\quad
    |\mathbf{\hat{2}}^I\rangle=|\mathbf{2}^I\rangle-z|1\rangle\frac{\langle\mathbf{2}^I1\rangle}{E+p}.
  \end{equation}

  The shifted spinors can also be written as the linear combinations of spinors from $p_1$ and $\bar{p}_1$,
  \begin{equation}
    \begin{aligned}
      |\hat{1}]&=|1]-z|\mathbf{2}^1]\frac{\langle\mathbf{2}^2 1\rangle}{E+p}=|1]+z|\bar{1}],\\
      |\mathbf{\hat{2}}^2\rangle&=|\mathbf{2}^2\rangle-z|1\rangle\frac{\langle\mathbf{2}^2 1\rangle}{E+p}=-\sqrt{\frac{E+p}{2p}}|\bar{1}\rangle+z\sqrt{\frac{2p}{E+p}}|1\rangle,\\
      |\mathbf{\hat{2}}^1\rangle&=|\mathbf{2}^1\rangle-z|1\rangle\frac{\langle\mathbf{2}^1 1\rangle}{E+p}=-\sqrt{\frac{E-p}{2p}}|1\rangle,\\
    \end{aligned}
  \end{equation}
  where we used $|\mathbf{2}^I]\langle\mathbf{2}_I|=-|\mathbf{2}^1]\langle\mathbf{2}^2|+|\mathbf{2}^2]\langle\mathbf{2}^1|$.

  Now the accurate expressions of the shifted momenta are
  \begin{equation}
    \begin{aligned}
      \hat{p}_1=p_1+zpr,\quad\hat{p}_2=p_2-zpr,
    \end{aligned}
  \end{equation}
  where the shift-vector is
  \begin{equation}
    \begin{aligned}
      r^\mu=\frac{\langle1|\gamma^\mu|\mathbf{2}^I]\langle\mathbf{2}_I1\rangle}{2p(E+p)}=-\frac{\langle1|\gamma^\mu|\mathbf{2}^1]\langle\mathbf{2}^2 1\rangle}{2p(E+p)}=\frac{\langle1|\gamma^\mu|\bar{1}]}{2p}=(0,-1,-i,0)
    \end{aligned}
  \end{equation}

  Since we want to discuss the large-$z$ behavior, the normalization of polarization vectors is unnecessary. The polarization vectors of particle 1 and 2 are
  \begin{equation}
    \begin{aligned}
      \epsilon^\mu_+(\hat{p}_1)&=\frac{\langle q|\gamma^\mu|\hat{1}]}{\langle q1\rangle},\quad
      \epsilon^\mu_-(\hat{p}_1)=\frac{\langle 1|\gamma^\mu|q]}{[\hat{1}q]},\\
      \epsilon^\mu_+(\hat{p}_2)&=-\frac{\langle \mathbf{\hat{2}}^1|\gamma^\mu|\mathbf{2}^1]}{m},\quad
      \epsilon^\mu_-(\hat{p}_2)=\frac{\langle\mathbf{\hat{2}}^2|\gamma^\mu|\mathbf{2}^2]}{m},\\
      \epsilon^\mu_0(\hat{p}_2)&=-\frac{\langle\mathbf{\hat{2}}^1|\gamma^\mu|\mathbf{2}^2]+\langle\mathbf{\hat{2}}^2|\gamma^\mu|\mathbf{2}^1]}{2m}.\\
    \end{aligned}
  \end{equation}

  We choose the reference vector $q=\bar{p}_1$, and these polarization vectors reduce to
  \begin{equation}
    \begin{aligned}
      \epsilon^\mu_+(\hat{p}_1)&=\frac{\langle \bar{1}|\gamma^\mu|1]}{\langle \bar{1}1\rangle}+z\frac{\langle \bar{1}|\gamma^\mu|\bar{1}]}{\langle \bar{1}1\rangle}=\frac{\langle \bar{1}|\gamma^\mu|1]}{2p}+z\frac{\langle \bar{1}|\gamma^\mu|\bar{1}]}{2p}=r^{*\mu}+z\frac{\bar{p}_1^\mu}{p},\\
      \epsilon^\mu_-(\hat{p}_1)&=\frac{\langle 1|\gamma^\mu|\bar{1}]}{[1\bar{1}]-z[\bar{1}\bar{1}]}=\frac{\langle 1|\gamma^\mu|\bar{1}]}{2p}=r^\mu,\\
      \epsilon^\mu_+(\hat{p}_2)&=\frac{\langle 1|\gamma^\mu|\bar{1}]}{2p}=r^\mu,\\
      \epsilon^\mu_-(\hat{p}_2)&=\frac{\langle\bar{1}|\gamma^\mu|1]}{2p}-z\frac{\langle1|\gamma^\mu|1]}{E+p}=r^{*\mu}-z\frac{2p_1^\mu}{E+p},\\
      \epsilon^\mu_0(\hat{p}_2)&=\frac{(E+p)\langle\bar{1}|\gamma^\mu|\bar{1}]-(E-p)\langle1|\gamma^\mu|1]}{4mp}-z\frac{\langle1|\gamma^\mu|\bar{1}]}{2m}=\frac{\bar{p}_2^\mu}{m}-z\frac{p}{m}r^\mu,\\
    \end{aligned}
    \label{5polar}
  \end{equation}
  where $\bar{p}_2=(p,0,0,-E)$ characterizes the longitudinal polarization of particle 2. If we set $p=1$, eq.\,\eqref{5polar} becomes eqs. \eqref{polar1} and \eqref{polar2}.

\section{Diagrammatic expressions}
  \label{AppC}

  \setcounter{equation}{0}
  \renewcommand{\theequation}{C.\arabic{equation}}
  In section \ref{sec4}, we expand amplitude $A(z)$ at $z=\infty$. There are many diagrammatic expressions corresponding to different contributions in this expansion. Consider the amplitudes with vector bosons. Since we set particle 1 to have negative helicity, the leftmost vector in these diagrammatic expressions must be $p_1$.

  Consider $N^{(0)}$, which includes only 3-vertices. Suppose the rightmost vector is $k$, the leading contribution is
  \begin{equation}\begin{aligned}
    \parbox[b]{4cm}{\HardNum{p_1}{R}{R}{R}{k}{\mu_3}{\mu_4}{\mu_i}\vspace{-0.2cm}}=c\left[2^{i-2}(p_1 k)\prod_{j=3}^i r_{\mu_j}-\sum_{l=3}^i 2^{l-3}(r k)p_{1\mu_l}\prod_{j\neq l} r_{\mu_j}\right],
    \label{MostR}
  \end{aligned}\end{equation}
  where $c$ corresponds to the color factor. When $k=r$ or $p_1$, it vanishes and refers to the leading diagram. When $k=\bar{p}_2$ or $r^*$, it corresponds to a subleading diagram.

  For $N^{-,0}$ and $N^{-,+}$, the subleading diagrams will give the non-vanish terms. We take $N^{-,0}_{\mu_3\mu_4\mu_5}$ as an example. First consider the diagram with only 3-vertices
  \begin{equation}\begin{aligned}
    N^{(0),-,0}_{\mu_3\mu_4\mu_5}
    =&z^2\parbox[b]{0.3cm}{\Bigg(\vspace{-0.2cm}}\parbox[b]{3.18cm}{\HardNumexample{p_1}{V}{R}{R}{r}{\mu_3}{\mu_4}{\mu_5}\vspace{-0.2cm}}+
    \parbox[b]{3.18cm}{\HardNumexample{p_1}{R}{V}{R}{r}{\mu_3}{\mu_4}{\mu_5}\vspace{-0.2cm}}+
    \parbox[b]{3.18cm}{\HardNumexample{p_1}{R}{R}{V}{r}{\mu_3}{\mu_4}{\mu_5}\vspace{-0.2cm}}\\
    &\ \ \ -\parbox[b]{3.3cm}{\HardNumexample{p_1}{R}{R}{R}{\bar{p}_2}{\mu_3}{\mu_4}{\mu_5}\vspace{-0.2cm}}\parbox[b]{0.3cm}{\Bigg)\vspace{-0.2cm}}+\mathcal{O}(z).
  \end{aligned}\end{equation}
  Evaluating these diagrammatic expressions, we get
  \begin{equation}\begin{aligned}
      N^{(0),-,0}_{\mu_3\mu_4\mu_5}
      =&c_1[-p_{1\mu_3}r_{\mu_4}r_{\mu_5}(5(p_3r)+3(p_4r)+(p_5r))\\
      &-2r_{\mu_3}p_{1\mu_4}r_{\mu_5}((p_3r)+3(p_4r)+(p_5r))\\
      &+4r_{\mu_3}r_{\mu_4}p_{1\mu_5}((p_3r)+(p_4r)-(p_5r))].\\
  \end{aligned}\end{equation}
  Then we consider the diagram with one 4-vertex.
  \begin{equation} \begin{aligned}
    N^{(1),3,-,0}_{\mu_3\mu_4\mu_5}=z\parbox[b]{2.5cm}{\begin{tikzpicture}
    \draw[thick]
        (0,0)node[circle,inner sep=0pt,draw=white,fill=white]{$p_1$}
      --(1*0.7,0)node[circle,inner sep=0pt,draw=white,fill=white]{$V$}
      --(2*0.7,0)node[circle,inner sep=0pt,draw=white,fill=white]{$R$}
      --(3*0.7,0)node[circle,inner sep=0pt,draw=white,fill=white]{$r$};
    \draw[thick]
        (1*0.7-0.3,0.7)node[circle,inner sep=0pt,draw=white,fill=white]{$\mu_3$}
      --(1*0.7,0)node[circle,inner sep=0pt,draw=white,fill=white]{$V$}
      --(1*0.7+0.25,0.7)node[circle,inner sep=0pt,draw=white,fill=white]{$\mu_4$};
    \draw[thick]
        (2*0.7,0)node[circle,inner sep=0pt,draw=white,fill=white]{$R$}
      --(2*0.7,0.7)node[circle,inner sep=-1pt,draw=white,fill=white]{$\mu_5$};
    \end{tikzpicture}\vspace{-0.2cm}}+\mathcal{O}(1),
  \end{aligned} \end{equation}
  \begin{equation} \begin{aligned}
    N^{(1),4,-,0}_{\mu_3\mu_4\mu_5}=z\parbox[b]{2.5cm}{\begin{tikzpicture}
    \draw[thick]
        (0,0)node[circle,inner sep=0pt,draw=white,fill=white]{$p_1$}
      --(1*0.7,0)node[circle,inner sep=0pt,draw=white,fill=white]{$V$}
      --(2*0.7,0)node[circle,inner sep=0pt,draw=white,fill=white]{$R$}
      --(3*0.7,0)node[circle,inner sep=0pt,draw=white,fill=white]{$r$};
    \draw[thick]
        (1*0.7,0)node[circle,inner sep=0pt,draw=white,fill=white]{$R$}
      --(1*0.7,0.7)node[circle,inner sep=-1pt,draw=white,fill=white]{$\mu_3$};
    \draw[thick]
        (2*0.7-0.25,0.7)node[circle,inner sep=0pt,draw=white,fill=white]{$\mu_4$}
      --(2*0.7,0)node[circle,inner sep=0pt,draw=white,fill=white]{$V$}
      --(2*0.7+0.3,0.7)node[circle,inner sep=0pt,draw=white,fill=white]{$\mu_5$};
    \end{tikzpicture}\vspace{-0.2cm}}+\mathcal{O}(1).
  \end{aligned} \end{equation}
  Evaluating these diagrammatic expressions, we get
  \begin{equation}\begin{aligned}
      N^{(1),3,-,0}_{\mu_3\mu_4\mu_5}=2(c_1-c_3)r_{\mu_3}r_{\mu_4}p_{1\mu_5}+2(c_3+c_2)r_{\mu_3}p_{1\mu_4}r_{\mu_5}+(-c_2-c_1)p_{1\mu_3}r_{\mu_4}r_{\mu_5},
  \end{aligned}\end{equation}
  \begin{equation}\begin{aligned}
      N^{(1),4,-,0}_{\mu_3\mu_4\mu_5}=(c_1-c_3)r_{\mu_3}p_{1\mu_4}r_{\mu_5}+(c_3+c_2)p_{1\mu_3}r_{\mu_4}r_{\mu_5}.
  \end{aligned}\end{equation}

  We do the decomposition and give the components
  \begin{equation}\begin{aligned}
      N^{3,-,0}_{\mu_3\mu_4\mu_5}=\frac{1}{2}(c_1-c_2+c_3)(p_{1\mu_3}r_{\mu_4}-r_{\mu_3}p_{1\mu_4})r_{\mu_5},
  \end{aligned}\end{equation}
  \begin{equation}\begin{aligned}
      N^{4,-,0,}_{\mu_3\mu_4\mu_5}=\frac{1}{2}(c_1-c_2+c_3)r_{\mu_3}(p_{1\mu_4}r_{\mu_5}-r_{\mu_4}p_{1\mu_5}).
  \end{aligned}\end{equation}

\end{fmffile}

\bibliographystyle{JHEP}
\bibliography{massiverecursion}

\providecommand{\href}[2]{#2}\begingroup\raggedright\begin{thebibliography}{10}

\bibitem{Britto:2004ap}
R.~Britto, F.~Cachazo and B.~Feng, \emph{{New recursion relations for tree
  amplitudes of gluons}},
  \href{https://doi.org/10.1016/j.nuclphysb.2005.02.030}{\emph{Nucl. Phys. B}
  {\bfseries 715} (2005) 499}
  [\href{https://arxiv.org/abs/hep-th/0412308}{{\ttfamily hep-th/0412308}}].

\bibitem{Britto:2005fq}
R.~Britto, F.~Cachazo, B.~Feng and E.~Witten, \emph{{Direct proof of tree-level
  recursion relation in Yang-Mills theory}},
  \href{https://doi.org/10.1103/PhysRevLett.94.181602}{\emph{Phys. Rev. Lett.}
  {\bfseries 94} (2005) 181602}
  [\href{https://arxiv.org/abs/hep-th/0501052}{{\ttfamily hep-th/0501052}}].

\bibitem{Badger:2005zh}
S.D.~Badger, E.W.N.~Glover, V.V.~Khoze and P.~Svrcek, \emph{{Recursion
  relations for gauge theory amplitudes with massive particles}},
  \href{https://doi.org/10.1088/1126-6708/2005/07/025}{\emph{JHEP} {\bfseries
  07} (2005) 025} [\href{https://arxiv.org/abs/hep-th/0504159}{{\ttfamily
  hep-th/0504159}}].

\bibitem{Risager:2005vk}
K.~Risager, \emph{{A Direct proof of the CSW rules}},
  \href{https://doi.org/10.1088/1126-6708/2005/12/003}{\emph{JHEP} {\bfseries
  12} (2005) 003} [\href{https://arxiv.org/abs/hep-th/0508206}{{\ttfamily
  hep-th/0508206}}].

\bibitem{Cachazo:2005ca}
F.~Cachazo and P.~Svrcek, \emph{{Tree level recursion relations in general
  relativity}},  \href{https://arxiv.org/abs/hep-th/0502160}{{\ttfamily
  hep-th/0502160}}.

\bibitem{Brandhuber:2008pf}
A.~Brandhuber, P.~Heslop and G.~Travaglini, \emph{{A Note on dual
  superconformal symmetry of the N=4 super Yang-Mills S-matrix}},
  \href{https://doi.org/10.1103/PhysRevD.78.125005}{\emph{Phys. Rev. D}
  {\bfseries 78} (2008) 125005}
  [\href{https://arxiv.org/abs/0807.4097}{{\ttfamily 0807.4097}}].

\bibitem{Arkani-Hamed:2008owk}
N.~Arkani-Hamed, F.~Cachazo and J.~Kaplan, \emph{{What is the Simplest Quantum
  Field Theory?}}, \href{https://doi.org/10.1007/JHEP09(2010)016}{\emph{JHEP}
  {\bfseries 09} (2010) 016} [\href{https://arxiv.org/abs/0808.1446}{{\ttfamily
  0808.1446}}].

\bibitem{Cheung:2015ota}
C.~Cheung, K.~Kampf, J.~Novotny, C.-H.~Shen and J.~Trnka, \emph{{On-Shell
  Recursion Relations for Effective Field Theories}},
  \href{https://doi.org/10.1103/PhysRevLett.116.041601}{\emph{Phys. Rev. Lett.}
  {\bfseries 116} (2016) 041601}
  [\href{https://arxiv.org/abs/1509.03309}{{\ttfamily 1509.03309}}].

\bibitem{Cohen:2010mi}
T.~Cohen, H.~Elvang and M.~Kiermaier, \emph{{On-shell constructibility of tree
  amplitudes in general field theories}},
  \href{https://doi.org/10.1007/JHEP04(2011)053}{\emph{JHEP} {\bfseries 04}
  (2011) 053} [\href{https://arxiv.org/abs/1010.0257}{{\ttfamily 1010.0257}}].

\bibitem{Cheung:2015cba}
C.~Cheung, C.-H.~Shen and J.~Trnka, \emph{{Simple Recursion Relations for
  General Field Theories}},
  \href{https://doi.org/10.1007/JHEP06(2015)118}{\emph{JHEP} {\bfseries 06}
  (2015) 118} [\href{https://arxiv.org/abs/1502.05057}{{\ttfamily
  1502.05057}}].

\bibitem{Schwinn:2005pi}
C.~Schwinn and S.~Weinzierl, \emph{{Scalar diagrammatic rules for Born
  amplitudes in QCD}},
  \href{https://doi.org/10.1088/1126-6708/2005/05/006}{\emph{JHEP} {\bfseries
  05} (2005) 006} [\href{https://arxiv.org/abs/hep-th/0503015}{{\ttfamily
  hep-th/0503015}}].

\bibitem{Schwinn:2006ca}
C.~Schwinn and S.~Weinzierl, \emph{{SUSY ward identities for multi-gluon
  helicity amplitudes with massive quarks}},
  \href{https://doi.org/10.1088/1126-6708/2006/03/030}{\emph{JHEP} {\bfseries
  03} (2006) 030} [\href{https://arxiv.org/abs/hep-th/0602012}{{\ttfamily
  hep-th/0602012}}].

\bibitem{Schwinn:2007ee}
C.~Schwinn and S.~Weinzierl, \emph{{On-shell recursion relations for all Born
  QCD amplitudes}},
  \href{https://doi.org/10.1088/1126-6708/2007/04/072}{\emph{JHEP} {\bfseries
  04} (2007) 072} [\href{https://arxiv.org/abs/hep-ph/0703021}{{\ttfamily
  hep-ph/0703021}}].

\bibitem{Craig:2011ws}
N.~Craig, H.~Elvang, M.~Kiermaier and T.~Slatyer, \emph{{Massive amplitudes on
  the Coulomb branch of N=4 SYM}},
  \href{https://doi.org/10.1007/JHEP12(2011)097}{\emph{JHEP} {\bfseries 12}
  (2011) 097} [\href{https://arxiv.org/abs/1104.2050}{{\ttfamily 1104.2050}}].

\bibitem{Boels:2011zz}
R.H.~Boels and C.~Schwinn, \emph{{On-shell supersymmetry for massive
  multiplets}}, \href{https://doi.org/10.1103/PhysRevD.84.065006}{\emph{Phys.
  Rev. D} {\bfseries 84} (2011) 065006}
  [\href{https://arxiv.org/abs/1104.2280}{{\ttfamily 1104.2280}}].

\bibitem{Arkani-Hamed:2017jhn}
N.~Arkani-Hamed, T.-C.~Huang and Y.-t.~Huang, \emph{{Scattering Amplitudes For
  All Masses and Spins}},  \href{https://arxiv.org/abs/1709.04891}{{\ttfamily
  1709.04891}}.

\bibitem{Aoude:2019tzn}
R.~Aoude and C.S.~Machado, \emph{{The Rise of SMEFT On-shell Amplitudes}},
  \href{https://doi.org/10.1007/JHEP12(2019)058}{\emph{JHEP} {\bfseries 12}
  (2019) 058} [\href{https://arxiv.org/abs/1905.11433}{{\ttfamily
  1905.11433}}].

\bibitem{Ballav:2020ese}
S.~Ballav and A.~Manna, \emph{{Recursion relations for scattering amplitudes
  with massive particles}},
  \href{https://doi.org/10.1007/JHEP03(2021)295}{\emph{JHEP} {\bfseries 03}
  (2021) 295} [\href{https://arxiv.org/abs/2010.14139}{{\ttfamily
  2010.14139}}].

\bibitem{Herderschee:2019dmc}
A.~Herderschee, S.~Koren and T.~Trott, \emph{{Constructing $ \mathcal{N} $ = 4
  Coulomb branch superamplitudes}},
  \href{https://doi.org/10.1007/JHEP08(2019)107}{\emph{JHEP} {\bfseries 08}
  (2019) 107} [\href{https://arxiv.org/abs/1902.07205}{{\ttfamily
  1902.07205}}].

\bibitem{Franken:2019wqr}
R.~Franken and C.~Schwinn, \emph{{On-shell constructibility of Born amplitudes
  in spontaneously broken gauge theories}},
  \href{https://doi.org/10.1007/JHEP02(2020)073}{\emph{JHEP} {\bfseries 02}
  (2020) 073} [\href{https://arxiv.org/abs/1910.13407}{{\ttfamily
  1910.13407}}].

\bibitem{ArkaniHamed:2008yf}
N.~Arkani-Hamed and J.~Kaplan, \emph{{On Tree Amplitudes in Gauge Theory and
  Gravity}}, \href{https://doi.org/10.1088/1126-6708/2008/04/076}{\emph{JHEP}
  {\bfseries 04} (2008) 076} [\href{https://arxiv.org/abs/0801.2385}{{\ttfamily
  0801.2385}}].

\bibitem{Cheung:2008dn}
C.~Cheung, \emph{{On-Shell Recursion Relations for Generic Theories}},
  \href{https://doi.org/10.1007/JHEP03(2010)098}{\emph{JHEP} {\bfseries 03}
  (2010) 098} [\href{https://arxiv.org/abs/0808.0504}{{\ttfamily 0808.0504}}].

\bibitem{Britto:2021pud}
R.~Britto, R.~Gonzo and G.R.~Jehu, \emph{{Graviton particle statistics and
  coherent states from classical scattering amplitudes}},
  \href{https://arxiv.org/abs/2112.07036}{{\ttfamily 2112.07036}}.

\bibitem{Elvang:2013cua}
H.~Elvang and Y.-t.~Huang, \emph{{Scattering Amplitudes}},
  \href{https://arxiv.org/abs/1308.1697}{{\ttfamily 1308.1697}}.

\bibitem{Christensen:2018zcq}
N.~Christensen and B.~Field, \emph{{Constructive standard model}},
  \href{https://doi.org/10.1103/PhysRevD.98.016014}{\emph{Phys. Rev. D}
  {\bfseries 98} (2018) 016014}
  [\href{https://arxiv.org/abs/1802.00448}{{\ttfamily 1802.00448}}].

\bibitem{Gunion:1989we}
J.F.~Gunion, H.E.~Haber, G.L.~Kane and S.~Dawson, \emph{{The Higgs Hunter's
  Guide}}, vol.~80 (2000).

\bibitem{He:2017jjx}
S.-P.~He, Y.-n.~Mao, C.~Zhang and S.-h.~Zhu, \emph{{$ZH\eta$ vertex in the
  simplest little Higgs model}},
  \href{https://doi.org/10.1103/PhysRevD.97.075005}{\emph{Phys. Rev. D}
  {\bfseries 97} (2018) 075005}
  [\href{https://arxiv.org/abs/1709.08929}{{\ttfamily 1709.08929}}].

\bibitem{Lee:1977eg}
B.W.~Lee, C.~Quigg and H.~Thacker, \emph{{Weak Interactions at Very
  High-Energies: The Role of the Higgs Boson Mass}},
  \href{https://doi.org/10.1103/PhysRevD.16.1519}{\emph{Phys. Rev. D}
  {\bfseries 16} (1977) 1519}.

\bibitem{Chanowitz:1985hj}
M.S.~Chanowitz and M.K.~Gaillard, \emph{{The TeV Physics of Strongly
  Interacting W's and Z's}},
  \href{https://doi.org/10.1016/0550-3213(85)90580-2}{\emph{Nucl. Phys. B}
  {\bfseries 261} (1985) 379}.

\bibitem{Ballav:2021ahg}
S.~Ballav and A.~Manna, \emph{{Recursion relations for scattering amplitudes
  with massive particles II: massive vector bosons}},
  \href{https://arxiv.org/abs/2109.06546}{{\ttfamily 2109.06546}}.

\bibitem{Feng:2009ei}
B.~Feng, J.~Wang, Y.~Wang and Z.~Zhang, \emph{{BCFW Recursion Relation with
  Nonzero Boundary Contribution}},
  \href{https://doi.org/10.1007/JHEP01(2010)019}{\emph{JHEP} {\bfseries 01}
  (2010) 019} [\href{https://arxiv.org/abs/0911.0301}{{\ttfamily 0911.0301}}].

\bibitem{Feng:2010ku}
B.~Feng and C.-Y.~Liu, \emph{{A Note on the boundary contribution with bad
  deformation in gauge theory}},
  \href{https://doi.org/10.1007/JHEP07(2010)093}{\emph{JHEP} {\bfseries 07}
  (2010) 093} [\href{https://arxiv.org/abs/1004.1282}{{\ttfamily 1004.1282}}].

\bibitem{Elvang:2016qvq}
H.~Elvang, C.R.T.~Jones and S.G.~Naculich, \emph{{Soft Photon and Graviton
  Theorems in Effective Field Theory}},
  \href{https://doi.org/10.1103/PhysRevLett.118.231601}{\emph{Phys. Rev. Lett.}
  {\bfseries 118} (2017) 231601}
  [\href{https://arxiv.org/abs/1611.07534}{{\ttfamily 1611.07534}}].

\bibitem{Falkowski:2020aso}
A.~Falkowski and C.S.~Machado, \emph{{Soft Matters, or the Recursions with
  Massive Spinors}}, \href{https://doi.org/10.1007/JHEP05(2021)238}{\emph{JHEP}
  {\bfseries 05} (2021) 238}
  [\href{https://arxiv.org/abs/2005.08981}{{\ttfamily 2005.08981}}].

\end{thebibliography}\endgroup

\end{document}